\documentclass[reprint,showpacs,showkeys,amsmath,amssymb,aps,prmaterials]{revtex4-2}

\usepackage{graphicx} 
\usepackage{rotating}
\usepackage{color}
\usepackage{float}
\usepackage{multirow}

\begin{document}

\preprint{Fecher; MnPtGa \today}

\title{Magnetic order and anisotropy in hexagonal MnPtGa \\
       {\normalsize -- Non-collinear magnetism contra frustration --} }

\author{Gerhard H. Fecher}
\email{fecher@cpfs.mpg.de}
\affiliation{Max-Planck-Institute for Chemical Physics of Solids,
             D-01187 Dresden, Germany.}

\author{Roshnee Sahoo}
\altaffiliation[Present address: ]{Binayak Acharya Degree College,
                Berhampur 760006, Odisha, India.}
\affiliation{Max-Planck-Institute for Chemical Physics of Solids,
             D-01187 Dresden, Germany.}

\author{Claudia Felser}
\email{felser@cpfs.mpg.de}
\affiliation{Max-Planck-Institute for Chemical Physics of Solids,
             D-01187 Dresden, Germany.}

\date{\today}

\begin{abstract}

MnPtGa is a hexagonal intermetallic compound with a rich variety of magnetic
orders. Its magnetic state is reported to range from collinear ferromagnetism, to
non-collinear skyrmion type order. MnPtGa is a system with strongly localized
magnetic moments at the Mn atoms as was demonstrated using calculations for
disordered local moments. The magnetic moments at the Mn sites stay at
$\approx 3.9\:\mu_B$ even above the calculated magnetic transition temperatures
($T_N=220$~K or $T_C=285$~K).

In the present work, a special emphasis was focused on the possible non-collinear
magnetic order using first principles calculations. The investigations
included magnetic anisotropy, static non-collinear orders in the form of spin
canting and dynamic non-collinearity in spin spirals. It is found that the
energy differences between ferromagnetic, antiferromagnetic, canted, or spiral
magnetic order are in the order of not more than 30~meV, which is in the order
of thermal energies at ambient temperature. This hints that a particular
magnetic state -- including skyrmions, antiskyrmions or spin glass transitions --
may be forced when an external field is applied at finite temperature.

\end{abstract}



\maketitle


\section{Introduction} 

Hexagonal, magnetic ternaries based on $d$ elements are interesting as they
might exhibit non-collinear spin structures as observed for example in the
hexagonal rare earth elements. Antiferromagnetism in the basal plane might lead
to frustration due to the triangular arrangement of the atoms. As a result, spin
glass type, spiral spin order, or other non-collinear magnetic order may appear.

MnPtGa, with a hexagonal crystal structure, was selected for the present study.
In an early report, it was unfortunately classified as a Heusler compound with
cubic $C1_b$ structure~\cite{HCr71}. Buschow and co-workers~\cite{BEn83, BMo84},
however, have shown that MnPtGa adopts a hexagonal Ni$_2$In derived crystal
structure with lattice parameters $a=4.328$~{\AA} and $c=5.576$~{\AA}.
Ferromagnetic order was reported with a magnetic moment of 3.15 to 4.72~$\mu_B$
per manganese atom and a Curie temperature in the order of 210 to
270~K~\cite{BEn83,BMo84,KSK86}. These early studies were all performed
for temperatures above 200~K, thus the low temperature magnetic structure stayed
rather unexplained. Cooley {\it et.~al}~\cite{CBS20} reported a variety of
different types of collinear and non-collinear magnetic orders while varying the
temperature between 4 and 300~K. The series of different magnetic order reported
in Reference~\cite{CBS20} is: collinear ferromagnetic at 200~K, canted
ferromagnet at 150~K, canted ferromagnet with spin density wave at below 50~K
with change of the period and increased intensity by further cooling to 10~K.
The pressure dependence of the magnetic order was discussed by Dubey
{\it et.~al}~\cite{DRJ24} for polycrystalline MnPtGa. Recently, measurements on
single-crystalline MnPtGa were reported and some aspects of the influence of
the non-collinear magnetic order on the band structure were explained by
Dwari {\it et.~al}~\cite{DDM24}.


Besides the above-mentioned non-collinear behavior, skyrmions were observed in
the Mn--Pt--Ga system at
temperatures below 220~K~\cite{SDS20}.
It was reported that the stability of the N{\'e}el--skyrmions depends on the
thickness of the investigated lamella. This was observed over a wide range of
temperatures. Further, a non-collinear magnetic order, with a canting of the
magnetic moments similar to the bulk material, was detected in epitaxial MnPtGa
thin films~\cite{ILO22,ILS22}. Non-Orthogonal spin currents were recently
investigated in MnPtGa films grown on MgO by Meng {\it et.~al}~\cite{MZC25}.


In the present work, collinear and non-collinear magnetic properties of MnPtGa
are investigated by means of first principles methods. Ferromagnetic and
antiferromagnetic interactions are investigated as well as properties related to
canted or spiral spin order.

\section{Details of the calculations} 

The electronic and magnetic structures of MnPtGa were calculated by means of
{\scshape Wien}2k~\cite{BSS90,BSM01,SBl02} and
{\scshape Sprkkr}~\cite{Ebe99,EKM11,KKR20,MEb22}
in the generalized gradient approximation (GGA). In particular,
the GGA functional in the parametrization of Perdew, Burke, and
Ernzerhof~\cite{PBE96} was used.
A test involving GGA+U resulted in magnetic moments that were
too high compared with experimental results.
{\scshape Wien}2k was used for the structural relaxation (volume and $c/a$) and
to test the magneto-crystalline anisotropy.
(Note that the forces on the atoms are identical Zero, due to
the symmetry of their positions. Therefore, their relative positions are not
changed during relaxation of the structure).
All other properties were calculated using
{\scshape Sprkkr}~\cite{EKM11}. Magnetic spin and orbital moments, magnetic
anisotropy, exchange coupling coefficients, Dzyaloshinskii--Moriya vectors,
transition temperatures, disordered local moments, and canted spin order were
all determined in the fully relativistic mode,
solving the spin polarized Dirac equation for core and
valence electrons.

The spin polarized semi-relativistic mode
(Dirac equation only for core electrons)
was used for spin spirals, due to symmetry
constraints~\cite{San91a,San91b}.
To visualize the electronic structure, the Bloch spectral
functions ($E(k,\rho)$)~\cite{FSt80} were calculated using the fully
relativistic mode. $E(k,\rho)$ can be interpreted as a kind of momentum ($k$)
resolved density of states ($\rho$). It is similar to the band structure.
The mean-field approach of Liechtenstein was used to determine
the magnetic transition temperatures~\cite{LKA87}. For multi-sublattice systems,
including antiferromagnets, Anderson's approach is used~\cite{And63}. Due to the
long-range interaction and the slow convergence of the exchange coupling
coefficients with cluster size, the spin wave stiffness constants and the
spiralisation coefficients can be determined using the extrapolation scheme of
Pajda {\it al}~\cite{PKT01,BSP24}.
More details of the calculations are found in References~\cite{KFF07b,FCF13} as
well as in additional citations given in the following sections.

\begin{figure}[htb]
   \centering
   \includegraphics[width=8cm]{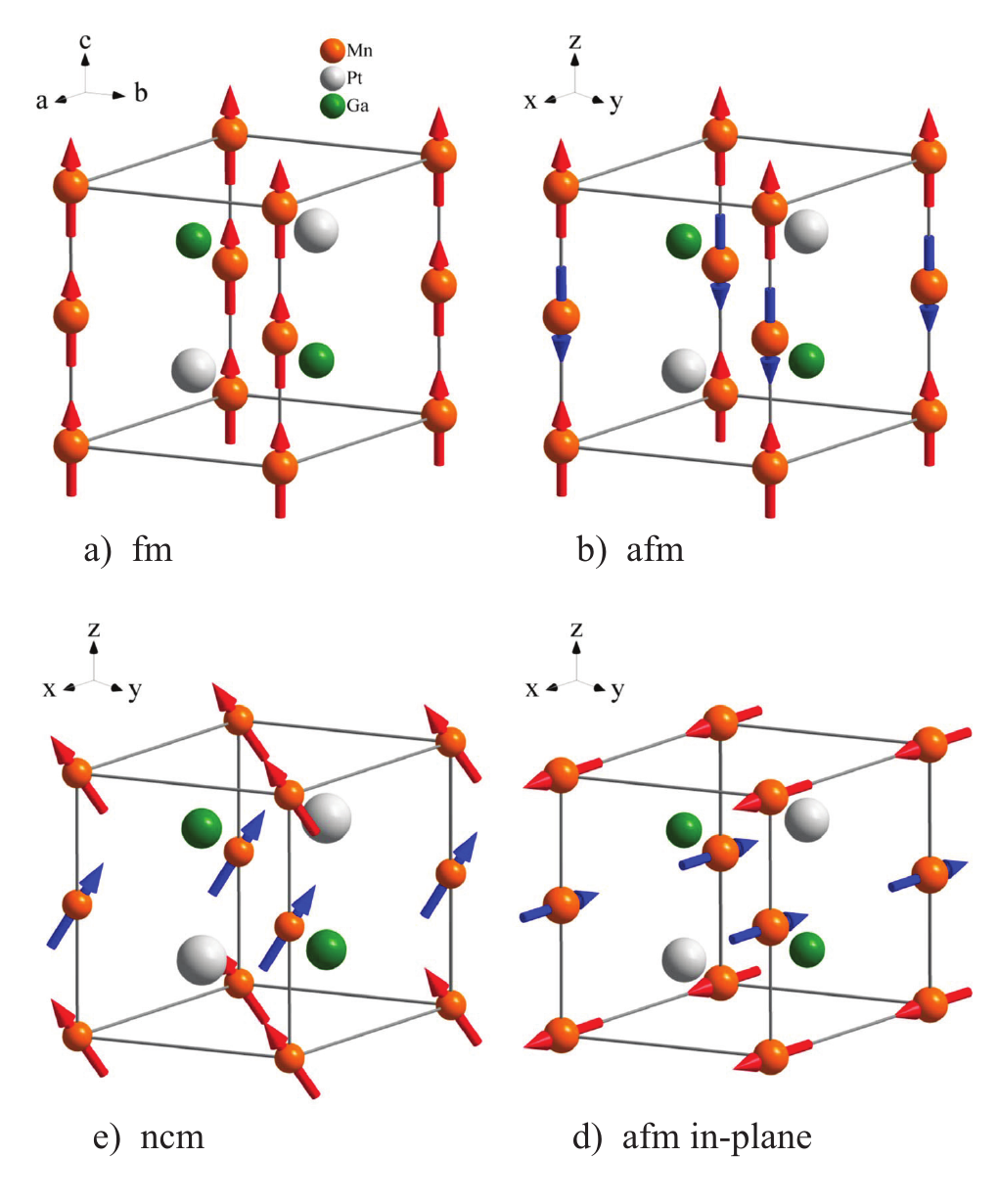}
   \caption{Crystal and magnetic structures of MnPtGa. \newline
            a) ferromagnetic, b) antiferromagnetic
            c) non collinear, d) antiferromagnetic in-plane.
            The arrows are drawn in the direction of the local magnetic moments
            at the Mn atoms.
            The angle between the magnetic moment vectors and the
            $z$ axis is in the canted, non-collinear state (c) is
            $\theta=35^\circ$ which is close to $\arctan(a/c)\approx38^\circ$.
            }
   \label{fig:struct}
\end{figure}

The basic crystal structure of MnPtGa (prototype: BeZrSi; $hP6$; $P\:6_3/mmc$
(194) $cad$) is shown in Figure~\ref{fig:struct} for
different magnetic structures.
The atoms are placed in the paramagnetic or
ferromagnetic structure on the 2a, 2c, and 2d Wyckoff positions of the hexagonal
cell. For calculation of the antiferromagnetic structure the Mn position is
split, the lattice symmetry is reduced to $P\:\overline{3}m1$ (164) where the
Mn$_\uparrow$, Mn$_\downarrow$, and Pt, Ga atoms are placed on 1a, 1b, and 2d
(with different position parameters for Pt and Ga), respectively.

The magnetic order changes the symmetry~\cite{Jos91}.
Depending on the direction of magnetization, incompatible
symmetry elements must be changed or removed. This results in the magnetic
(colored) space groups listed in Reference~\cite{CTP21}. The magnetic symmetry
is not necessarily the same as the lattice symmetry, but it is generally lower.
The magnetic space groups for various directions of magnetization can be
determined using {\scshape findsym}~\cite{SHa05,SHC05}.
The  magnetic space group is $P\:6_3/mm'c'$ (194.270) for the collinear
ferromagnetic order with spin magnetic moments $m_s$ along the $c$ axis. For the
antiferromagnetic order it becomes $P\:6'_3/m'm'c$ (194.268) when $m_s||c$.
Here, the prime $'$ is the spin reversal operator~\cite{Jos91} (sometimes called
time reversal operator). $m'$ is a mirror operation followed by spin reversal.
It should be mentioned that the colored group is $C\:m'c'm$ (63.462) or $C\:mcm$
(63.457) for the antiferromagnetic order with magnetic moments lying in the
basal plane. However, more complicated checker board type antiferromagnetic
order may also appear. For the later investigated non-collinear spin structure
the magnetic space group becomes $P\:2_1/m.1$ (11.50) or $C\:m'c'm$ (63.462).
In the simple antiferromagnetic or non-collinear cases the Wyckoff positions are
different, but the lattice parameters are kept hexagonal.

\section{Results and discussion} 

\subsection{Collinear magnetic structures.} 

The lattice parameters reported from experiments vary and are in the ranges
$(4.328\leq a \leq 4.446)$~{\AA} and
$(5.572 < c \leq 5.590)$~{\AA}~\cite{BEn83,BMo84,KSK86,CBS20,SDS20}.
Therefore, calculations of the electronic structure and magnetic properties were
performed for optimized lattice parameters. As starting point, the lattice
parameters of the two collinear magnetic structures were optimized using
{\scshape Wien}2k. Spin-orbit interaction was accounted for to respect the high
$Z$ of Pt and the spin-orbit splitting of the Ga $3d$ semi-core level. The
results of the optimization for collinear spin structures are summarized in
Table~\ref{tab:opt}. The calculated lattice parameters and positions of the
atoms fit very well to those reported from
experiments~\cite{BEn83,BMo84,KSK86,CBS20}.

\begin{table}[htb]
\centering
    \caption{Properties of MnPtGa with ferromagnetic (fm) or
             antiferromagnetic (afm), collinear magnetic order. \\
             Tabulated are the lattice parameters ($a$, $c$, $c/a$),
						 the Bulk modulus $B$ for hydrostatic pressure (fixed optimized $c/a$ ratio),
             the total energy ($E_{tot}$), and the magnetic moment
						 at the Mn atom $m_{\rm Mn}$ from the optimization with {\scshape Wien}2k.
             Experimental lattice parameters are from
             References~\cite{BEn83,BMo84,KSK86,CBS20,SDS20}. Experimental
             magnetic moments from~\cite{BEn83,CBS20} depend on the
             temperature. }
    \begin{ruledtabular}
    \begin{tabular}{l ccc}
                              & fm      &  afm     & experiment \\
       \hline
       $a$   [\AA]            & 4.3552  &  4.3571  & $4.328 \ldots 4.446$ \\
       $c$   [\AA]            & 5.5306  &  5.5318  & $5.572 \ldots 5.590$ \\
       $c/a$                  & 1.2699  &  1.2696  & $1.257 \ldots 1.289$ \\
       \hline
			 $B$ [GPa]              & 112     &  151     &                      \\
       $E_{tot}$ [Ry]         & -86198.140678 & -86198.143094 &           \\
			 \hline
       $m_{\rm Mn}$ [$\mu_B$] & 3.7232  &  $\pm 3.7105$ & $2.63 \ldots 3.15$ \\
    \end{tabular}
    \end{ruledtabular}
    \label{tab:opt}
\end{table}

\begin{figure}[htb]
   \centering
   \includegraphics[width=8cm]{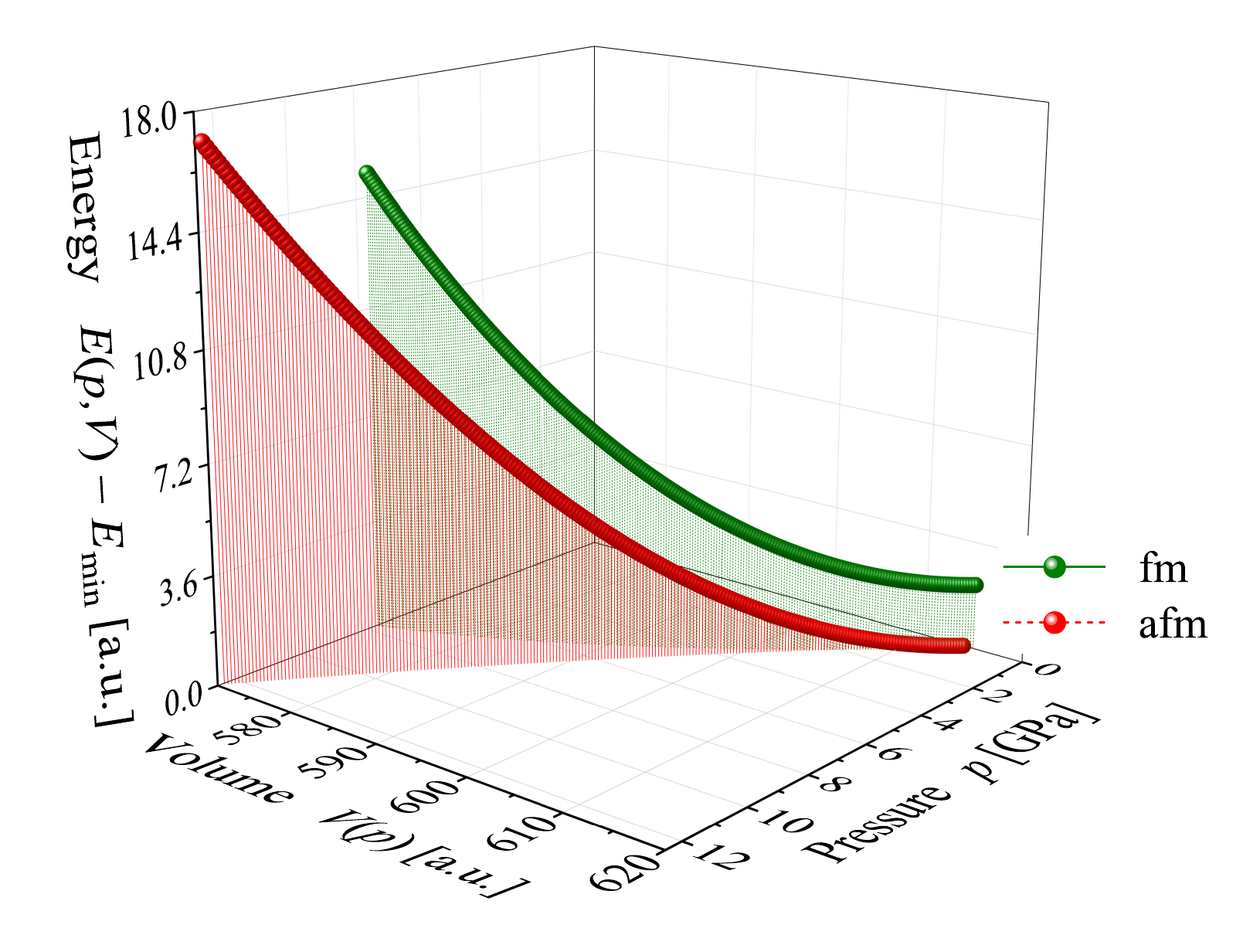}
   \caption{Calculated equation of state for MnPtGa with ferromagnetic (fm) or
            antiferromagnetic (afm), collinear magnetic order. \\
            Shown are the Energy--Volume--Pressure relations ($E(V,p)-E_{\min}$)
            according to the Birch--Murnaghan equation of
            state~\cite{Mur44,Bir47}. $E_{\min}$ is the equilibrium energy of
            the antiferromagnetic state (see Table~\ref{tab:opt}). Magnetization
            is along [0001].
            (Note the use of atomic units for energy (mRy) and volume
            ($a_{0B}^3$, $a_{0B}=0.529\ldots$~{\AA})).}

   \label{fig:vol_eos}
\end{figure}

The equilibrium lattice parameters of the ferromagnetic and antiferromagnetic
states are very close together. For a better comparison, the
Birch--Murnaghan~\cite{Mur44,Bir47}
equation of state {--energy versus volume and pressure--} curves for the two
collinear magnetic orderings are shown in Figure~\ref{fig:vol_eos}. The
equilibrium energies at ambient pressure ($p=0$) differ by only 33~meV, that is
in the order of thermal energies at room temperature. The lowest energy is
obtained for the antiferromagnetic state (see also Table~\ref{tab:opt}). This
gives a hint that the low temperature magnetic state may differ from the high
temperature ferromagnetic state observed in experiments.

The magnetic moment at the Mn atoms is nearly independent of the magnetic order.
The small difference is caused by the slightly different lattice parameters. Its
value of $\approx3.7\:\mu_B$ as well as the ferromagnetic Curie temperature are
in agreement with the findings from experiments~\cite{BEn83,BMo84,KSK86}. The
calculated Curie temperature is 65~K higher compared to the N{\`e}el
temperature. Therefore, it can be concluded that the high and low temperature
magnetic order is different, as already mentioned above.

\begin{figure}[htb]
   \centering
   \includegraphics[width=8cm]{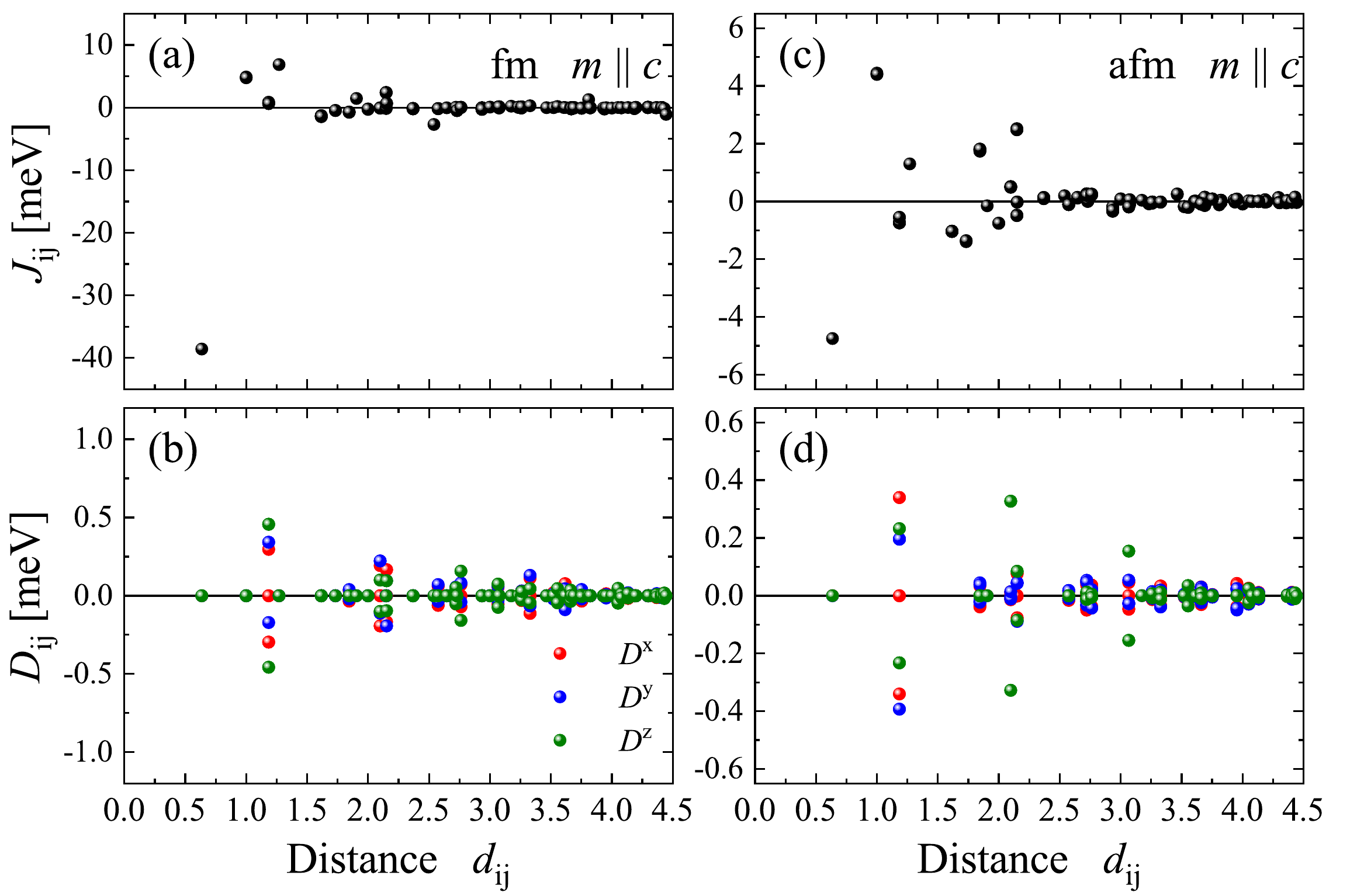}
   \caption{Exchange coupling coefficients of MnPtGa with ferromagnetic (fm) or
            antiferromagnetic (afm) order. \\
            Compared are the exchange coupling coefficients $J_{ij}$ in (a, c)
            and the Dzyaloshinskii--Moriya coefficients $D_{ij}$ in (b, d). The
            distance $d_{ij}=|r_{ij}|/a$ between the Mn atoms is given relative
            to the lattice parameter $a$. Magnetic moments are assumed to be
            aligned along the $c$ axis. (Note the different energy scales.)}
   \label{fig:exchange}
\end{figure}

The exchange coupling coefficients $J_{ij}$ needed to calculate the transition
temperatures~\cite{LKA87,PKT01,EMa09,MEb17} are plotted in
Figure~\ref{fig:exchange} for the case of ferromagnetic or antiferromagnetic
ordered collinear magnetic moments.
In micromagnetic or atomistic spin dynamics calculations,
magnetic order is ferromagnetic when the inter-sublattice exchange coupling
coefficients are positive, and antiferromagnetic when they are negative. In
electronic structure calculations, the individual $J_{ij}$  values can be either
positive or negative depending on the distance $r_{ij}$, regardless of the
magnetic order. The large negative values for the interaction between nearest
neighbors indicate that the ferromagnetic state is not stable, at least at low
temperatures.
Further shown are the antisymmetric exchange constants
$D^{x,y,z}_{ij}$~\cite{EMa09} that are the components of
the Dzyaloshinskii--Moriya interaction~\cite{Mor60,Dzy64} vectors. The $D_{ij}$
are considerably smaller than the $J_{ij}$, however, the latter drop down faster
with increasing distance between the Mn atoms.

The isotropic $J_{ij}$ and anisotropic $D_{ij}$ exchange coupling coefficients
form the exchange tensor:

\begin{equation}
  \mathbb{J}_{ij}=
  \begin{pmatrix}
     J_{ij}     &  D_{ij}^{z} & -D_{ij}^{y} \\
    -D_{ij}^{z} &  J_{ij}     &  D_{ij}^{x} \\
     D_{ij}^{y} & -D_{ij}^{x} &  J_{ij}     \\
  \end{pmatrix}.
\end{equation}

where $J_{ij}=\frac{1}{3}\sum_{\alpha}{J_{ij}^{\alpha\alpha}}$
and $D_{ij}^{\gamma}=\frac{1}{2}(J_{ij}^{\alpha\beta}-J_{ij}^{\beta\alpha}$)
($\alpha,\beta,\gamma=x,y,z$, $\alpha\neq\beta\neq\gamma$)~\cite{USP03,BKN21}.

Further, the exchange coupling coefficients may be used to
determine effective spin model parameters for micromagnetic calculations.
These are the spin wave stiffness ($A$) and spiralization ($D$):

\begin{eqnarray}
  A & = & \sum_{i \neq j} J_{ij} (r_{ij} \otimes r_{ij}) \nonumber \\
  D & = & \frac{1}{2}\sum_{i \neq j} D_{ij} \otimes r_{ij},
\end{eqnarray}

where $r_{ij}$ are the difference vectors between the positions
of atoms $i$ and $j$ ($i \neq j$) and $\otimes$ is the tensor
product~\cite{GHH19,BKN21}. The spin stiffness constant $A$ describes a
quadratic dependence of the spin wave energy on the wave vector $q$, that is,
$E_{\rm SW}=Aq^2$. It should not be confused with the exchange stiffness
constant.

The allowed combination of the vector components $D_{ij}^{\alpha}$ of the
exchange coupling coefficients and spatial coordinates $r_{ij}^{\beta}$ depend
on the investigated geometry ($\alpha,\beta$ assign the vector components
$x,y,z$ of $D_{ij}$ and $r_{ij}$).

The general components of the spiralization matrix are
$D^{\gamma}=\sum_{i\neq j}{D_{ij}^\alpha r_{ij}^\beta}$
For a magnetic monolayer (ML) with $C_{3v}$ ($m3$) point-group
symmetry the effective stiffness and spiralization are
$A_{\rm ML}=\frac{1}{4}\sum J_{ij}(d_{ij}^{y})^2$ and
$D_{\rm ML}=\sum D_{ij}^{x}\cdot d_{ij}^{y}$, respectively~\cite{SRP18} (here
$d_{ij}=r_{ij}/a$).

All $D_{eff}$ are Zero in the present case because the investigated bulk
structure shown in Figure~\ref{fig:struct} is centrosymmetric~\cite{Mor60}. The
Mn sites have $D_{3d}$ ($\overline{3}m$) point group symmetry.
It is usually expected that skyrmions require
non-centrosymmetric crystal structures in which the spiralization does not
vanish. However, here the bulk material is centrosymmetric, which would render
the observed appearance of skyrmions impossible. At the surface, however, the
inversion symmetry is broken~\cite{CLa98}, because an inversion would equate the
bulk to the vacuum. Similarly, inversion symmetry is broken at the interface
with a different material. This suggests that skyrmions could exist in thin
films despite the centrosymmetry of the bulk material. Due to geometric
restrictions in the summation, in thin films the spiralization may not vanish.
This is analogous to the dipole anisotropy, which can differ between bulk
materials and thin films.

Thus, the vanishing of the spiralization for bulk materials
does not contradict the observed existence of skyrmions. In thin films or
lamellas, the summations in the effective stiffness and spiralization are
spatially restricted in at least one direction, which can lead to a
non-vanishing $D_{eff}$. This is a direct consequence of broken inversion
symmetry at surfaces and interfaces. This possibly explains the experimental
observation by Srivastava {\it et.~al}~\cite{SDS20} that the size and stability
of the skyrmions depend on the thickness of the investigated lamella.

\begin{figure}[htb]
   \centering
   \includegraphics[width=8cm]{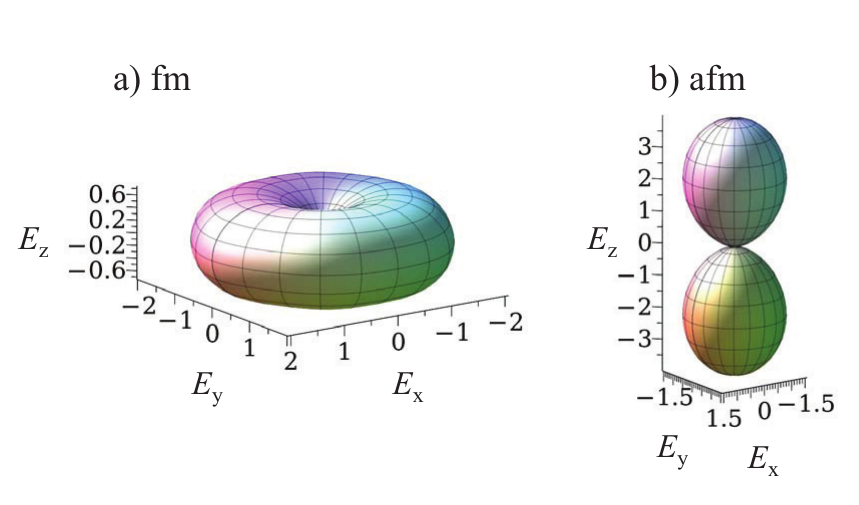}
   \caption{Magnetic anisotropy of ferromagnetic (fm) and antiferromagnetic (afm) MnPtGa.\\
            The magneto-crystalline anisotropy energy $E_{a'}$ is given in meV.
            In the ferromagnetic state, the {\it "easy"} direction is along $z$
            ($[0001]$), whereas three principle {\it "hard"} directions appear
            in the basal plane. The basal plane becomes an {\it "easy"} plane in
            the antiferromagnetic state. }
   \label{fig:aniso}
\end{figure}

Further, the directional dependence of the magnetization was investigated to
explain details of the collinear magnetic order. In particular, the total energy
was calculated for the cases where the magnetization is pointing along different
crystallographic directions. The obtained energy differences allow one to
determine the magneto-crystalline anisotropy (see also Appendix~\ref{app:app1}).
The result is visualized in Figure~\ref{fig:aniso}.

In the ferromagnetic state, the anisotropy is almost uniaxial with the {\it
"easy"} axis (lowest energy) along $z$ ($[0001]$) and an appreciable variation
in the basal plane. It is hardest along $x$, that is in the direction between to
nearest neighbors in the basal plane. The anisotropy behaves differently in the
antiferromagnetic state. Here the $z$-axis and is the {\it "hard"} axis (highest
energy)  pointing along $[0001]$. The basal plane becomes an easy plane. A
closer inspection reveals that there are three main easy axes found in the basal
plane that are along $x$ ($[21\overline{3}0]$) and similar. However, the energy
differences to the $y$-axis ($[01\overline{1}0]$) are very tiny only, thus that
already small changes in the lattice parameters, method and parameters of the
calculations are able to change the situation (see also
Reference~\cite{KBE16} for a comparison of different methods). The second order
uniaxial anisotropy constants are $K^{\rm fm}_u=2.85$~MJ/m$^3$ and
$K^{\rm afm}_u=-6.85$~MJ/m$^3$ for the ferromagnetic and antiferromagnetic
states, respectively. These values result in anisotropy fields of
$\mu_0 H^{\rm fm}_a=7.5$~T and $\mu_0 H^{\rm afm}_a=18.1$~T. From these
observations it is concluded that the ferromagnetic order prefers a nearly
uniaxial anisotropy with $c$ as easy axis whereas the result of
antiferromagnetic order is a triaxial anisotropy with the spins aligned in the
basal plane. In both cases $a$ and $b$ are hard axes. Care has to be taken if
the anisotropy energy is determined while applying a magnetic field which
changes the magnetic energy. The consequences are examined in
Appendix~\ref{app:app2}.

Further, the dipolar magneto-crystalline anisotropy was calculated from a direct
lattice sum of the dipolar energy (see Appendix~\ref{app:dipani}):

\begin{equation}
     E_{\rm dipaniso}=E_{dip}(\vec{n}_1)- E_{dip}(\vec{n}_2),
\label{eq:dip}
\end{equation}

where $\vec{n}=\vec{M}/M$ is the direction of magnetization. Here, the
anisotropy energy $\Delta E_{\rm dipaniso}$ is the difference of the dipolar
energies for the two different directions $\vec{n}_1=[0001]$ and
$\vec{n}_2=[1000]$. These are the directions along the $c$-axis and in the basal
plane along $a$. Using these directions in Equation~(\ref{eq:dip}), the dipolar
magnetic anisotropies for the collinear, ferromagnetic and antiferromagnetic
states are found to be
$\Delta E_{\rm dipaniso}^{\rm fm}=-74$~$\mu$eV (-0.13~MJ/m$^3$) and
$\Delta E_{\rm dipaniso}^{\rm afm}=119$~$\mu$eV (0.21~MJ/m$^3$), respectively.
These values are about 20 to 30 times lower compared to the magneto-crystalline
anisotropy energies. The positive value of the antiferromagnetic state points on
an easy dipolar direction that is in the basal plane, whereas the ferromagnetic
state has an easy dipolar axis along the $c$ axis.

The dipolar anisotropy is rather small compared to the anisotropy calculated
from the total energies. Here, it was calculated for a sphere with a radius of
50~nm. The situation will change for other shapes resulting in a distinct shape
anisotropy. In particular, in thin films the dimension perpendicular to the film
is much smaller compared to the dimensions in the film plane. Therefore, the
summation in Equation~(\ref{eq:dipaniso}) of Appendix~\ref{app:dipani} becomes a
truncated sphere that is strongly anisotropic and one will obtain a pronounced
thin film anisotropy. Additionally, this thin film anisotropy will be affected
by magnetic moments that differ at interfaces and surfaces from the center of
the film.

The results of the calculations with {\scshape Sprkkr} for the
magnetic properties of MnPtGa in the collinear state are summarized in
Table~\ref{tab:magprop2}.

\begin{table}[htb]
\centering
    \caption{Magnetic properties of MnPtGa. \\
             Tabulated are the total energies ($E_{tot}$), the magnetic moments at
             the Mn atoms ($m_{\rm Mn}$), the uniaxial crystalline anisotropies
             ($K_u$), the dipolar anisotropies ($\Delta E_{\rm dipaniso}$), and
             the magnetic transition temperature $T_m$, where $m=C$, or $N$ for
             the Curie or N{\`e}el temperature, respectively.
						 Calculations were performed with {\scshape Sprkkr}.
						 Experimental $T_m$ values are for ferromagnetic
             transitions~\cite{BEn83,BMo84,KSK86,CBS20,ILO22,ILS22}. The
             magnetic moments from~\cite{BEn83,CBS20} depend on the temperature.}
    \begin{ruledtabular}
    \begin{tabular}{l ccccc}
                                & fm   &  afm  & ncm  & experiment \\
       \hline
       $|m_{\rm Mn}|$ [$\mu_B$] & 3.72 &  3.71 & 3.18 & $2.63 \ldots 3.15$ \\
       $T_m$ [K]                & 285  &  220  &      & $210 \ldots 263$ \\
       $K_u$ [meV]              & 1.61 & -3.89 &      & & \\
       $\Delta E_{\rm dipaniso}$ [meV] & -0.07 & 0.12 & & \\
    \end{tabular}
    \end{ruledtabular}
    \label{tab:magprop2}
\end{table}

\subsection{Disordered local moments.}

\begin{figure}[htb]
   \centering
   \includegraphics[width=8cm]{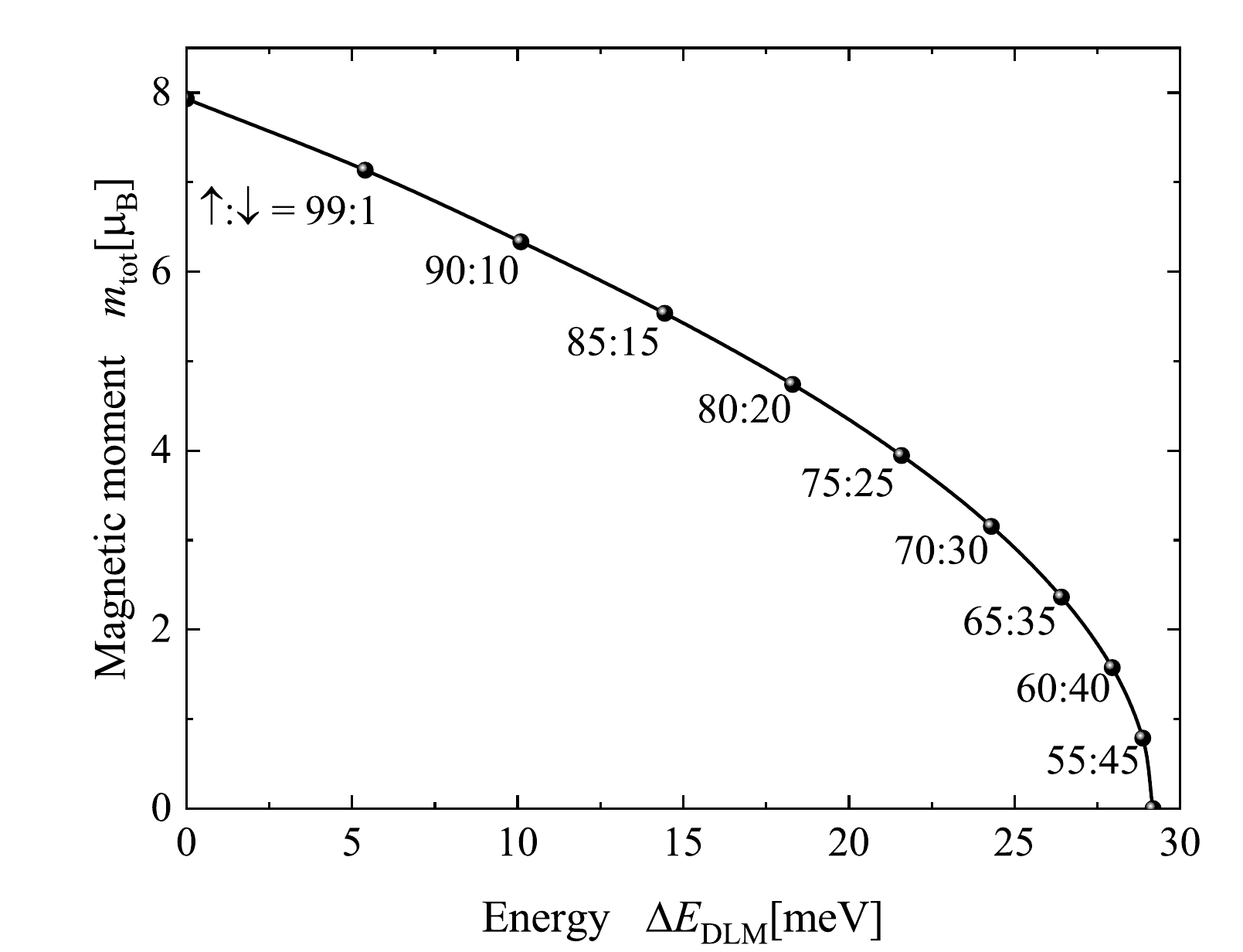}
   \caption{Disordered local moments in MnPtGa.\\
           The total (spin plus orbital) magnetic moment in the primitive cell
           containing 2 Mn atoms is plotted as function of the energy difference
           between the ferromagnetic and disordered local moment state.
           $\uparrow:\downarrow$ is the ratio of spin up to spin down electrons
           at the Mn site.}
   \label{fig:dlm-magn}
\end{figure}

Compounds containing Mn are typically systems with localized magnetic moments.
Here the coherent potential approximation (CPA)~\cite{Sov67,ADe93} was used to
model the random distribution of the magnetic moments at the Mn atoms, what is
known as the disordered local moment (DLM)~\cite{PSS83,SGP85} method. At each Mn
site, two species with opposite directions and equal sizes of the magnetic moments
were assumed starting from 99\% spin up ($\uparrow$) and 1\% spin down
($\downarrow$) up to 50\% of both spin directions. That is, the calculations
were performed for the pseudo-alloy (Mn$_{(1-x)\uparrow}$Mn$_{x\downarrow}$)PtGa.
It was found that the magnetic moment at the Mn sites stays in the order of
3.9~$\mu_B$ whereas the total magnetic moment of MnPtGa vanishes (see
Figure~\ref{fig:dlm-magn}). This is in good agreement with the experimental work
of Buschow and de~Mooij~\cite{BMo84}, who found an effective paramagnetic moment
of 3.6~$\mu_B$ per Mn atom from the slope of the inverse susceptibility
$\chi^{- 1}(T)$ curve at high temperature ($T>300$~K).

Figure~\ref{fig:dlm-magn} shows the total magnetic moments (spin+orbital) in the
primitive cell (containing two Mn atoms) as function of the energy difference
between the fully ferromagnetic and completely spin disordered states. The
energy difference for complete spin disorder amounts to 29.2~meV. This
corresponds to $\Delta E/k_B\approx170$~K per Mn atom.

\begin{figure}[htb]
   \centering
   \includegraphics[width=8cm]{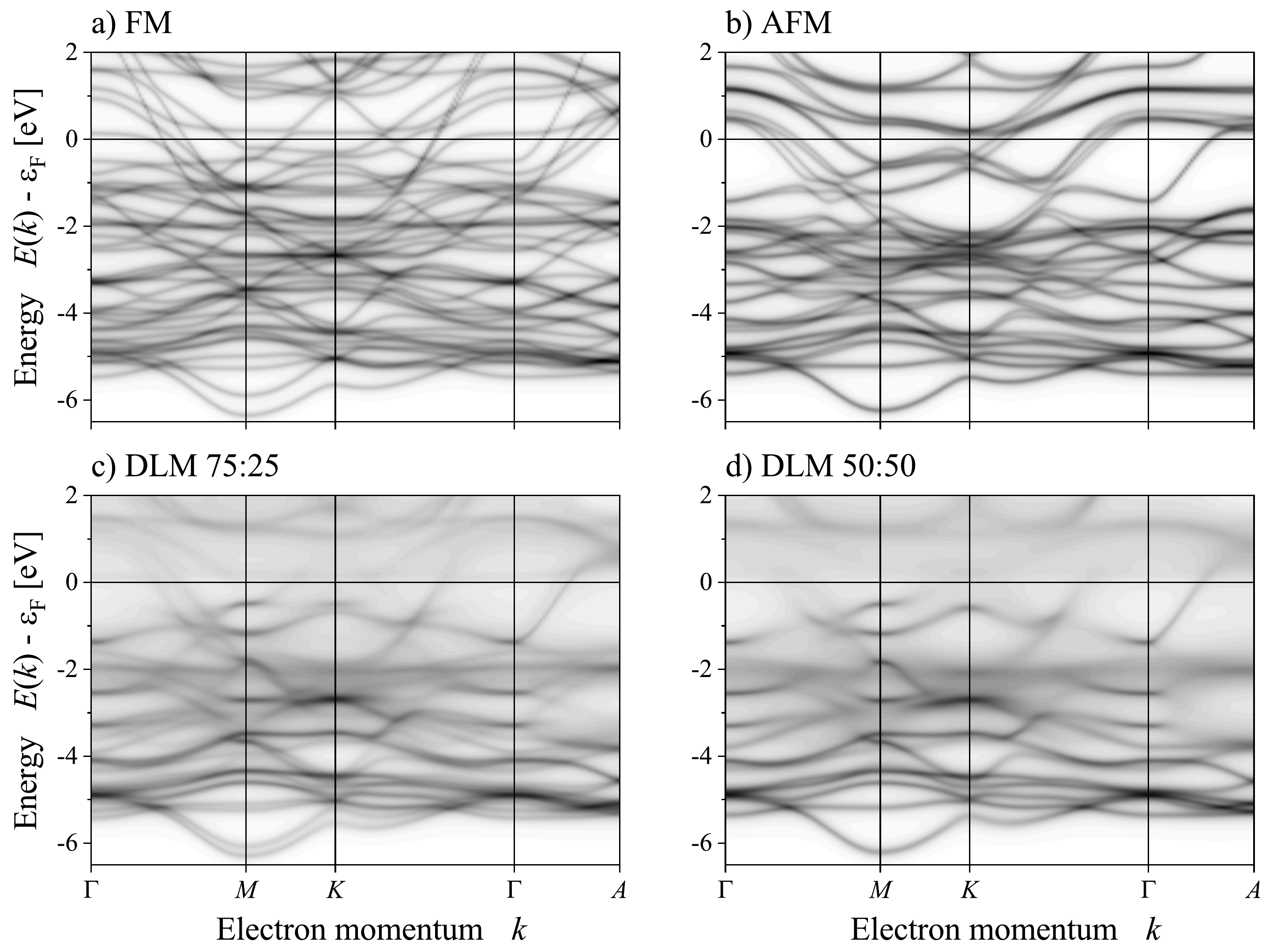}
   \caption{Electronic structure for ordered and disordered local moments in MnPtGa.\\
            The Bloch spectral functions of the ferromagnetic (a) and
            antiferromagnetic (b) states are compared to the disordered local
            moment states with 50\% (c) and full (d) spin disorder.}
   \label{fig:dlm-bsf}
\end{figure}

Figure~\ref{fig:dlm-bsf} compares the electronic structure of MnPtGa with
ordered and disordered local magnetic moments. In the disordered local moment
state, the electronic states in the region of the Fermi energy ($\epsilon_F$)
are strongly broadened compared to the case with ordered moments, whereas those
farer away from $\epsilon_F$ are less affected. Comparing to the ferromagnetic
state, it is found that the exchange splitting is smaller in the disordered
local moment state. This becomes obvious for the band minima at the $M$ point,
where the splitting finally vanishes for full spin disorder or in the
antiferromagnetic state.
It is not possible to directly compare Figure~\ref{fig:dlm-bsf}
to the experiments of Cooley {\it et~al}~\cite{CBS20}, because the magnetic
order changes with temperature and cannot directly be compared to the
calculations for the purely collinear ferromagnetic state. Furthermore, these
types of calculations cannot be performed with an applied magnetic field, which
is used in the experiments.

\subsection{Spiral spin order.} 

\begin{figure}[htb]
   \centering
   \includegraphics[width=8cm]{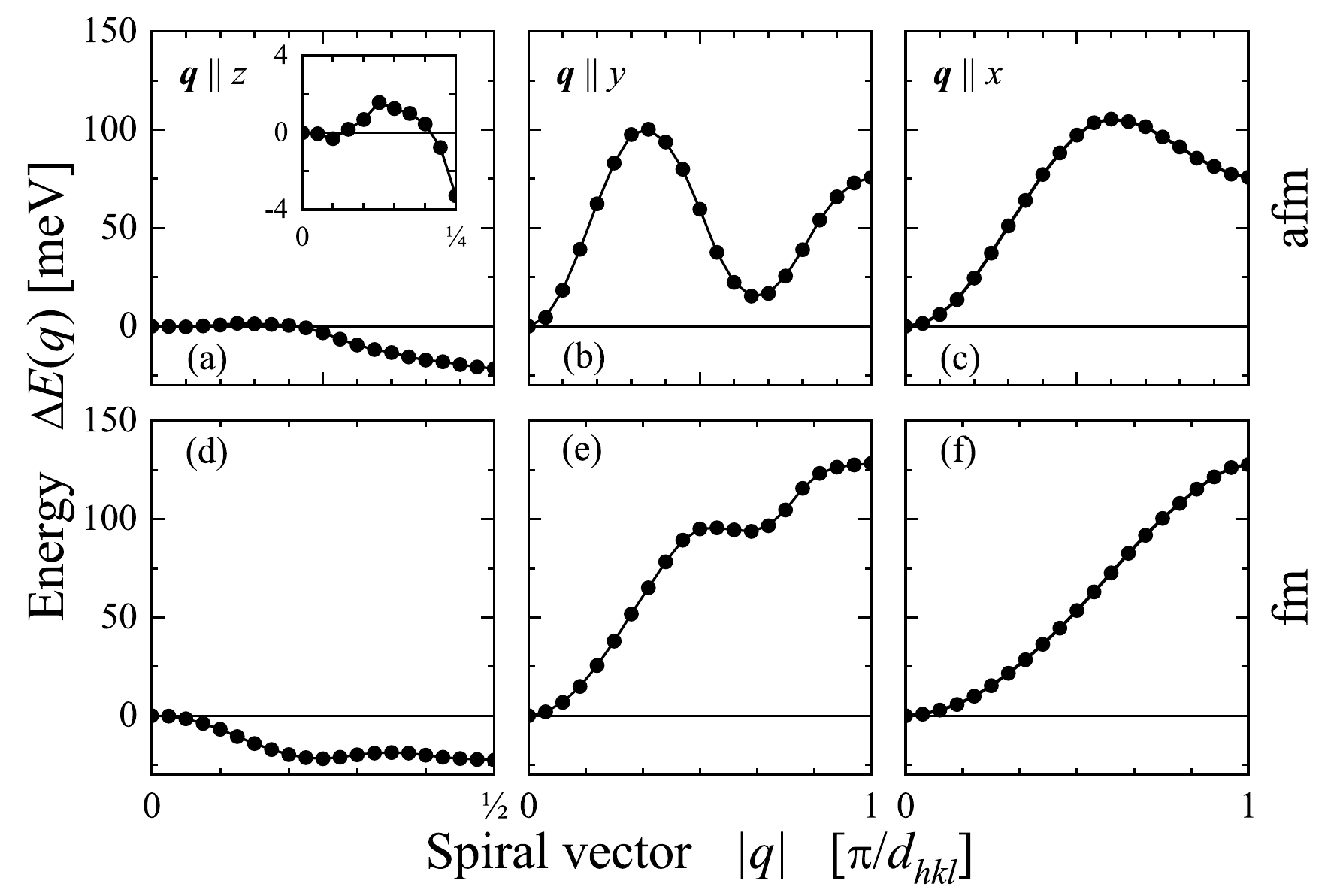}
   \caption{Planar spin spirals in antiferromagnetic (a-c) and ferromagnetic (d-f) MnPtGa.\\
            The spiral energies are given with respect to $q=0$, that is:
            $\Delta E(q)=E(q)-E(0)$. Note that the absolute energies at $q=0$
            differ between the starting spin order but are identical for
            different directions within a given type of magnetic order.}
   \label{fig:spiral_comp}
\end{figure}

The close energies and phase transition temperatures for the ferromagnetic and
antiferromagnetic states, together with the more complicated behavior of the
antiferromagnetic anisotropy, suggest that a non-collinear magnetic order appears
in MnPtGa. Therefore, spin spirals (see References~\cite{San91a,San91b,MFE11}
for details) were calculated for different directions and different starting
magnetic spin orders, either antiferromagnetic or ferromagnetic. In
centrosymmetric systems, the spin spirals are not influenced by spin--orbit or
Dzyaloshinskii--Moriya interaction~\cite{Mor60}. Note that the orientation of
the spins will not longer be the same as shown in Figure~\ref{fig:struct}, but
they will depend on the direction of propagation as well as the initial magnetic
order. In planar spirals, the spins are perpendicular to the propagation
direction. The energies are compared in Figure~\ref{fig:spiral_comp} for such
planar type spirals.

The spirals along $x$ or $y$ propagate in the hexagonal planes, where $y$ is in
the direction of a nearest neighbor atom (note: here, $y$ corresponds to
$[01\overline{1}0]$ in the hexagonal crystal co-ordinates). In these cases, the
lowest energy is observed at $q=0$ independent of the initial spin order. Most
interesting is the behavior along the $z$ axis ($[0001]$). For both initial
spin orders, the energy is clearly lower for $q=\pi/d_{0001}$ compared to 0. The
energy differences amount to about -20~meV. The energy is nearly constant in the
antiferromagnetic spiral up to $\approx0.5\pi/d_{0001}$. Similarly, it is observed
in the ferromagnetic spiral that the energy is nearly constant above
$\approx0.5\pi/d_{0001}$. Both might hint at frustrated states, as many spin
configurations will all have the same energy.
Further, when starting with
ferromagnetic spin order, the energy exhibits a local minimum at
$\approx0.5\pi/d_{0001}$. A discussion of the conditions for transitions between
collinear (e.g.: ferromagnet) or non-collinear (e.g.: spin spiral) arrangements
of the magnetic moments in elements was provided by
Liz{\'a}rraga {\it et~al}~~\cite{LNB04}. Planar spin spirals were also used to
explain the magnetic frustration, the spiral spin order as well as the
multidegenerate ground state of the rubidium sesquioxide
Rb$_4$O$_6$~\cite{WFJ09}.

\begin{figure}[htb]
   \centering
   \includegraphics[width=8cm]{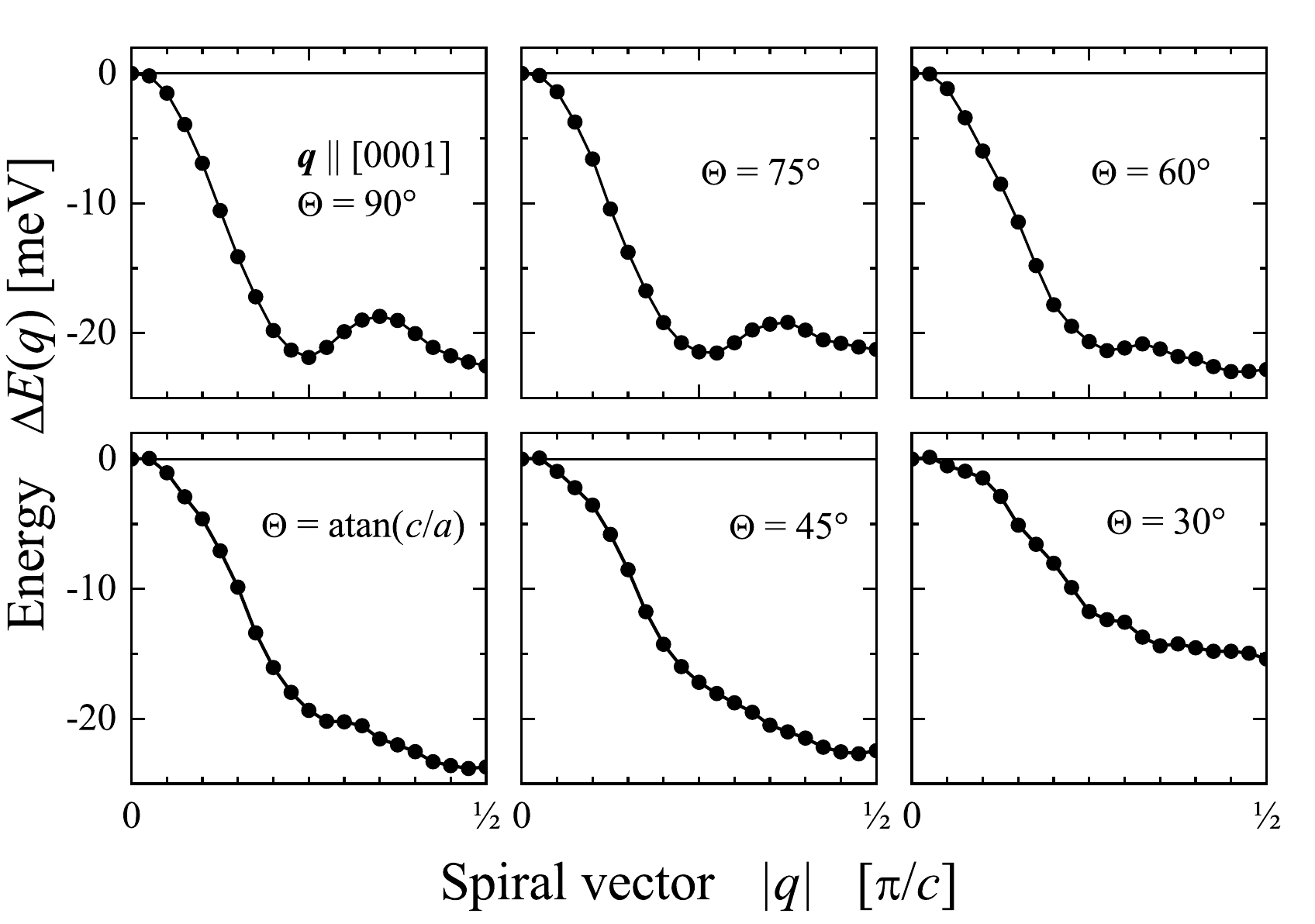}
   \caption{Energies of ferromagnetic cone spirals. }
   \label{fig:fm_cone-spiral}
\end{figure}

The spin direction was assumed to be perpendicular to the $q$ vector, in the
above calculations for planar spirals. Thus, the angle between $\vec{q}$ and
local magnetic moments $\vec{m}_i$ was set to $\Theta=\pi/2$. In a next step,
the spirals were assumed to be of cone type with $0<\Theta<\pi/2$, to allow for
a more detailed analysis. The calculations were performed for $q$ along
$[0001]$. The energies for $q$ in other directions are, of course, throughout
higher, as already observed for the planar spirals.
Figure~\ref{fig:fm_cone-spiral} displays the results for the ferromagnetic cone
spirals with variation of the cone angle. The lowest energy appears always
at $q=0.5\pi/c$. The local energy minimum observed in the planar spiral vanishes
with decreasing cone angle.
The ferromagnetic, cone type spin spirals have a negative
curvature at $q=0$, regardless of the propagation direction. Therefore, these
spiral structures exhibit negative spin stiffness constants. This corresponds to
the previously observed negative values of the exchange coupling coefficients
between nearest neighbors for collinear spins.

\begin{figure}[htb]
   \centering
   \includegraphics[width=8cm]{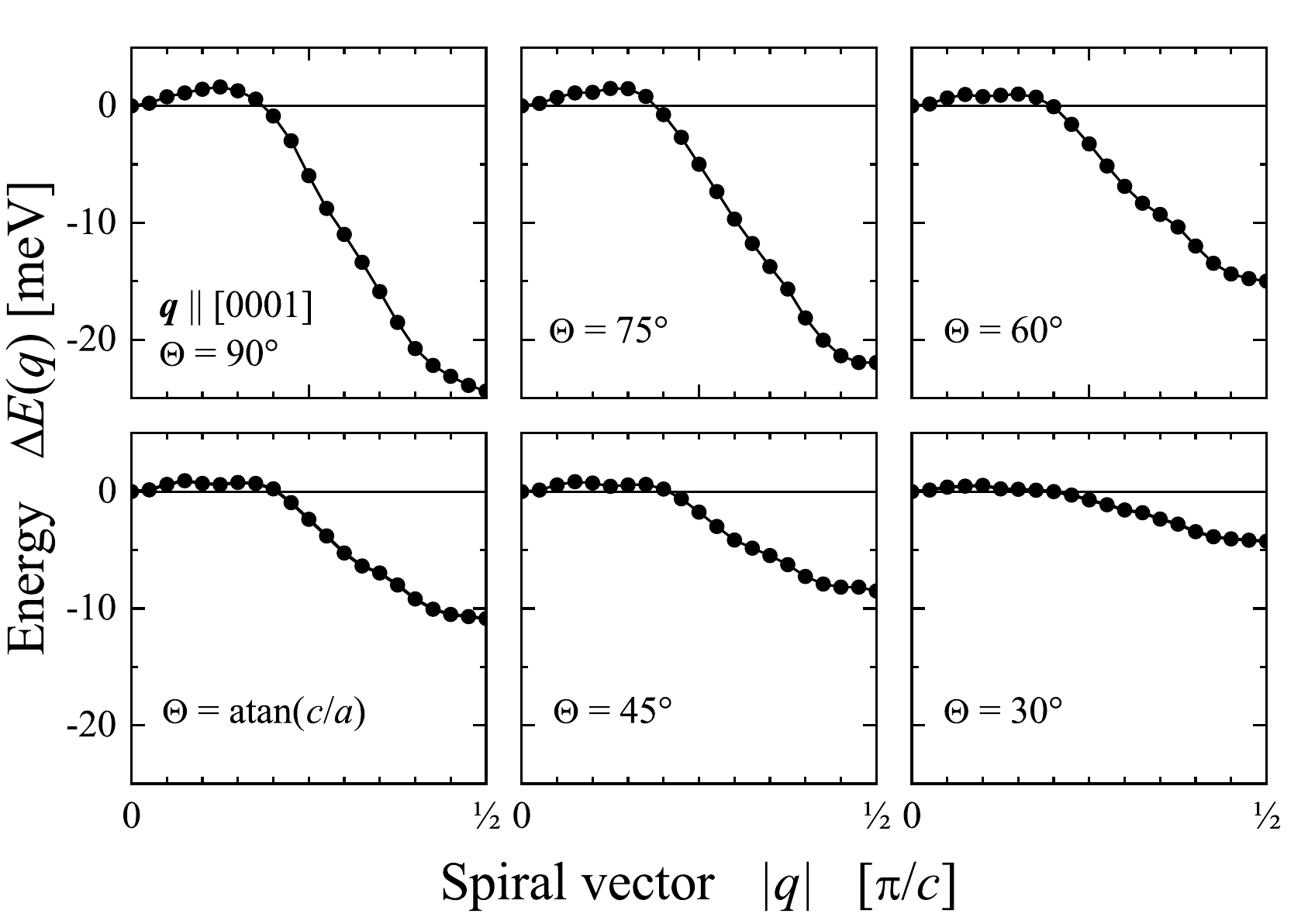}
   \caption{Energies of antiferromagnetic cone spirals. }
   \label{fig:afm_cone-spiral}
\end{figure}

The results for the antiferromagnetic cone spirals are displayed in
Figure~\ref{fig:afm_cone-spiral}. Similar to the ferromagnetic case, the lowest
energy appears always at $q=0.5\pi/c$. Starting from the planar spiral with
$\Theta=\pi/2$, the $q$ range with nearly unchanged energies increases slightly.

\begin{figure}[htb]
   \centering
   \includegraphics[height=8cm]{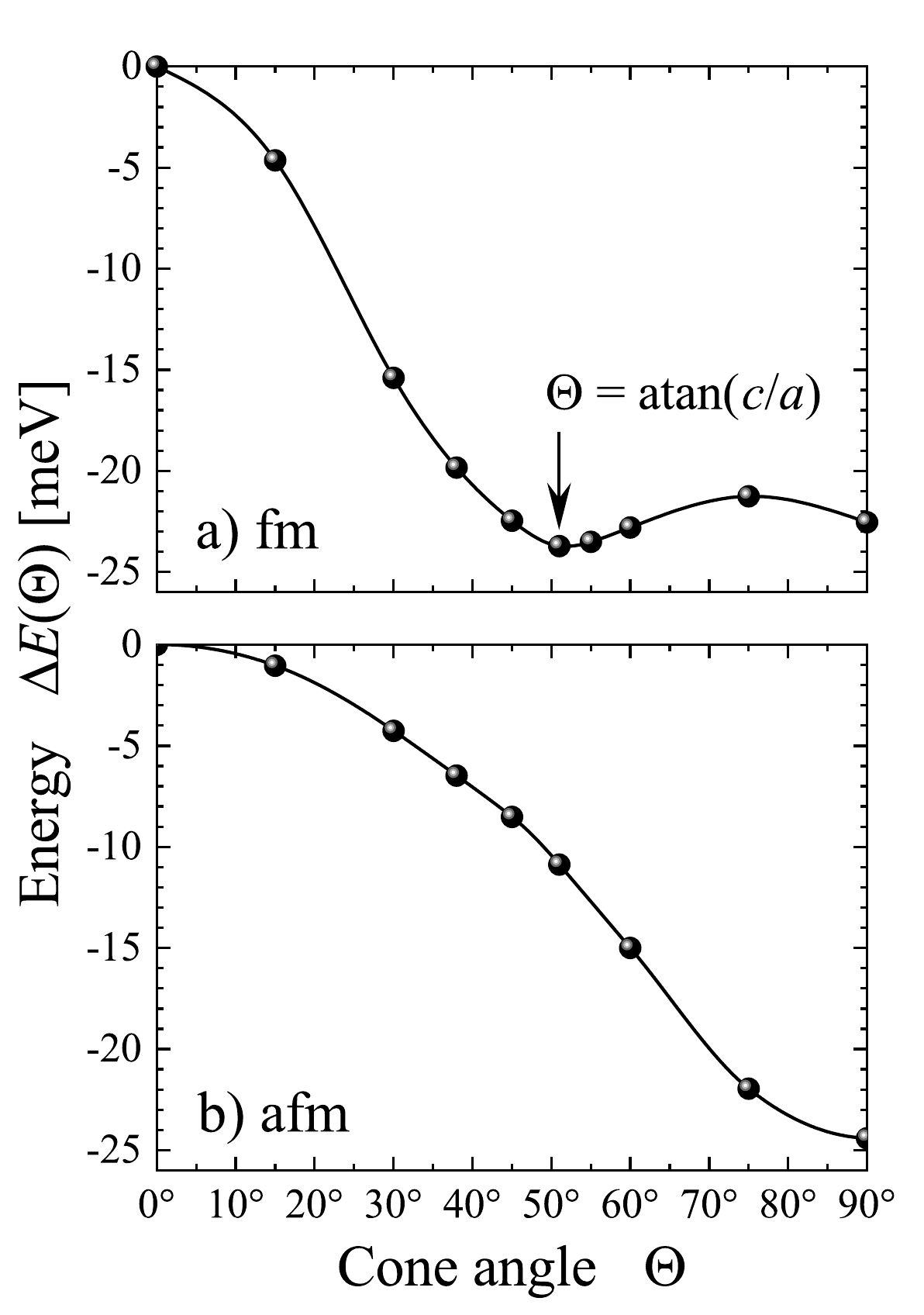}
   \caption{Minimum energies of the cone spirals. \\
            Compared are the minimum spiral energies at $q=0.5\pi/c$
            for initial ferromagnetic (fm) and
            antiferromagnetic (afm) spin order.}
   \label{fig:cone_comp}
\end{figure}

The dependence of the energy minimum on the cone angle $\Theta$ is compared in
Figure~\ref{fig:cone_comp}. The minimum energy of the antiferromagnetic cone
spiral decreases with decreasing cone angle. It is lowest for $\Theta=\pi/2$,
that is, for the planar spiral. The behavior of the ferromagnetic cone spiral is
different. After an initial decrease, only a weak variation is observed for
cone angles above $45^\circ$ and the lowest value appears already for
$\tan(\Theta) \approx c/a$ corresponding to approximately $51.78^\circ$. As a
result, one has a spiral with a ferromagnetic moment along its propagation
direction. The small energy variations at large $\Theta$ as well as large $q$
again hint on frustration, in particular in the basal plane of the lattice.

\subsection{Canted spin order.} 

\begin{figure}[htb]
   \centering
   \includegraphics[height=8cm]{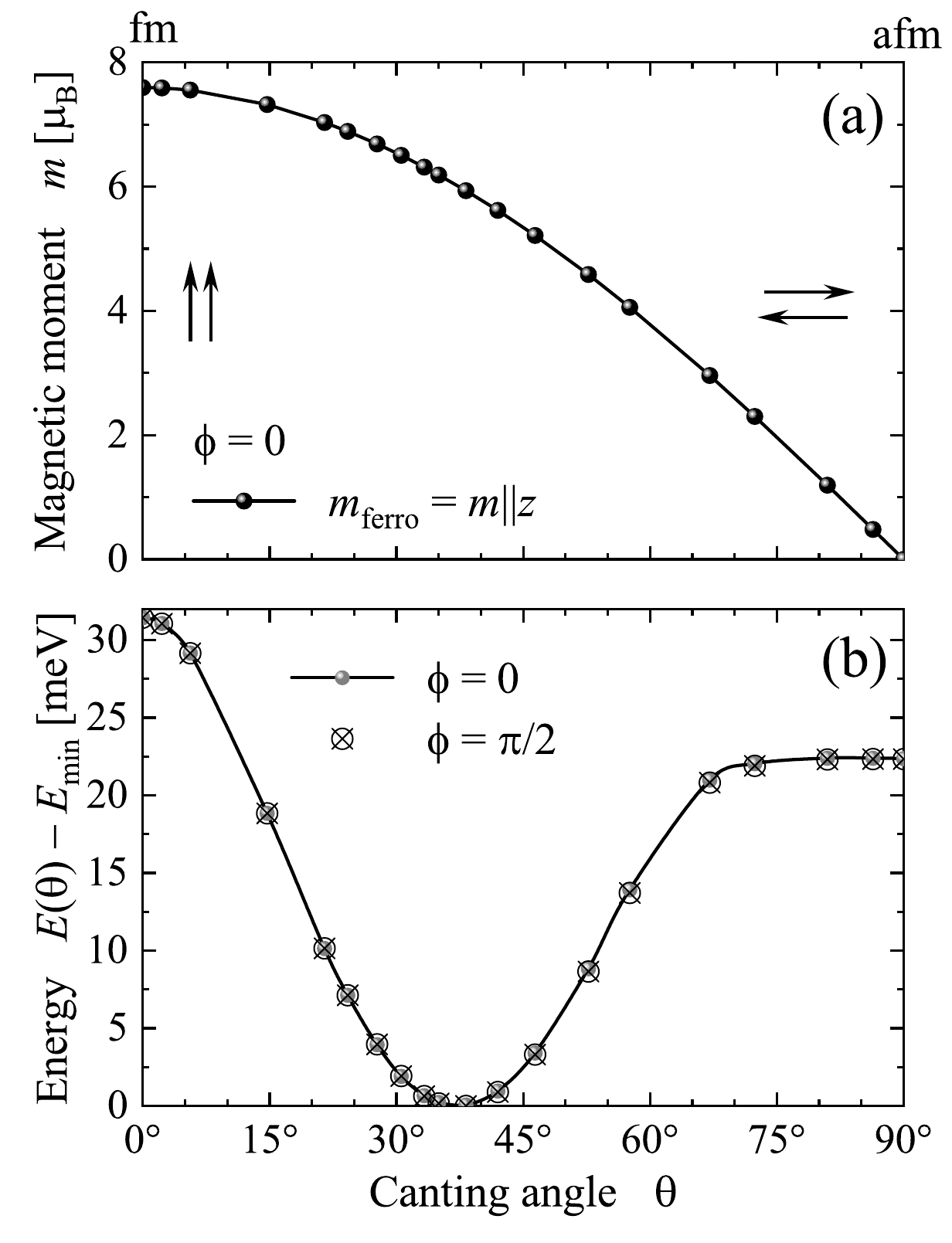}
   \caption{Magnetic moments and energies in the canted, non-collinear
            magnetic state of MnPtGa.\\
            The angle $\theta$ was varied with respect to the [0001] direction
            at fixed $\phi=0$ and $\phi=\pi/2$. The energy minimum appears at
            $\theta_{\min}=37.82^\circ \pm 0.15^\circ$.
            Note that the alignment of the magnetic moments is along the
            hexagonal $z$ axis for ferromagnetic type order, whereas it is in
            the basal plane for antiferromagnetic type order.}
   \label{fig:canted}
\end{figure}

In a last step, the energies were calculated for non-collinear, canted spin
orientation. It was assumed that the magnetic moments are aligned under a polar
angle $\theta$ with respect to the $c$ axis. The azimuthal angle $\phi$ is
measured with respect to the $x$ axis (corresponding to $[21\overline{3}0]$ in
crystal co-ordinates). Starting point is the anti-ferromagnetic structure with
two distinguished Mn atoms. The magnetic moments at the two Mn atoms were set to
have a ferromagnetic (parallel) orientation along the $c$ axis and an
antiferromagnetic (antiparallel) alignment in the basal plane by setting
$\phi_2=\phi_1+\pi$. It turned out that the lowest energies are obtained for
$\phi_{1,2}=0, \pi$. The resulting energies and magnetic moments are shown in
Figure~\ref{fig:canted}.

The ferromagnetic contribution to the spin magnetic moment varies between
$\theta=0$ and $\pi/2$ from 3.81~$\mu_B$ per primitive cell to 0, while going
from a ferromagnetic to an anti-ferromagnetic state. At the same time, the
energy exhibits a minimum at $\theta\approx38^\circ$ that is about 25 or 30~meV
lower compared to that of the antiferromagnetic or ferromagnetic states,
respectively. From this observation it is concluded that the magnetic ground
state of MnPtGa may be most probably canted. The corresponding
magneto-crystalline structure is shown in
Figure~\ref{fig:struct}(c).
(Note that the angle from Mn to a second nearest neighbor Pt atom is
$31.2^\circ$, and one finds, for example, that the magnetic moment is
approximately pointing from the Mn at (0,~1,~0) to the Pt atom at
(1/3,~2/3,~3/4).) The canting angle at the minimum of energy calculated here for
bulk material is larger compared to the one of $\approx 20^\circ$ determined at
2~K for thin films~\cite{ILO22}. The reason is that thin films have generally a
shape (or dipolar) anisotropy which is distinct from the bulk material.

\begin{figure}[htb]
   \centering
   \includegraphics[height=8cm]{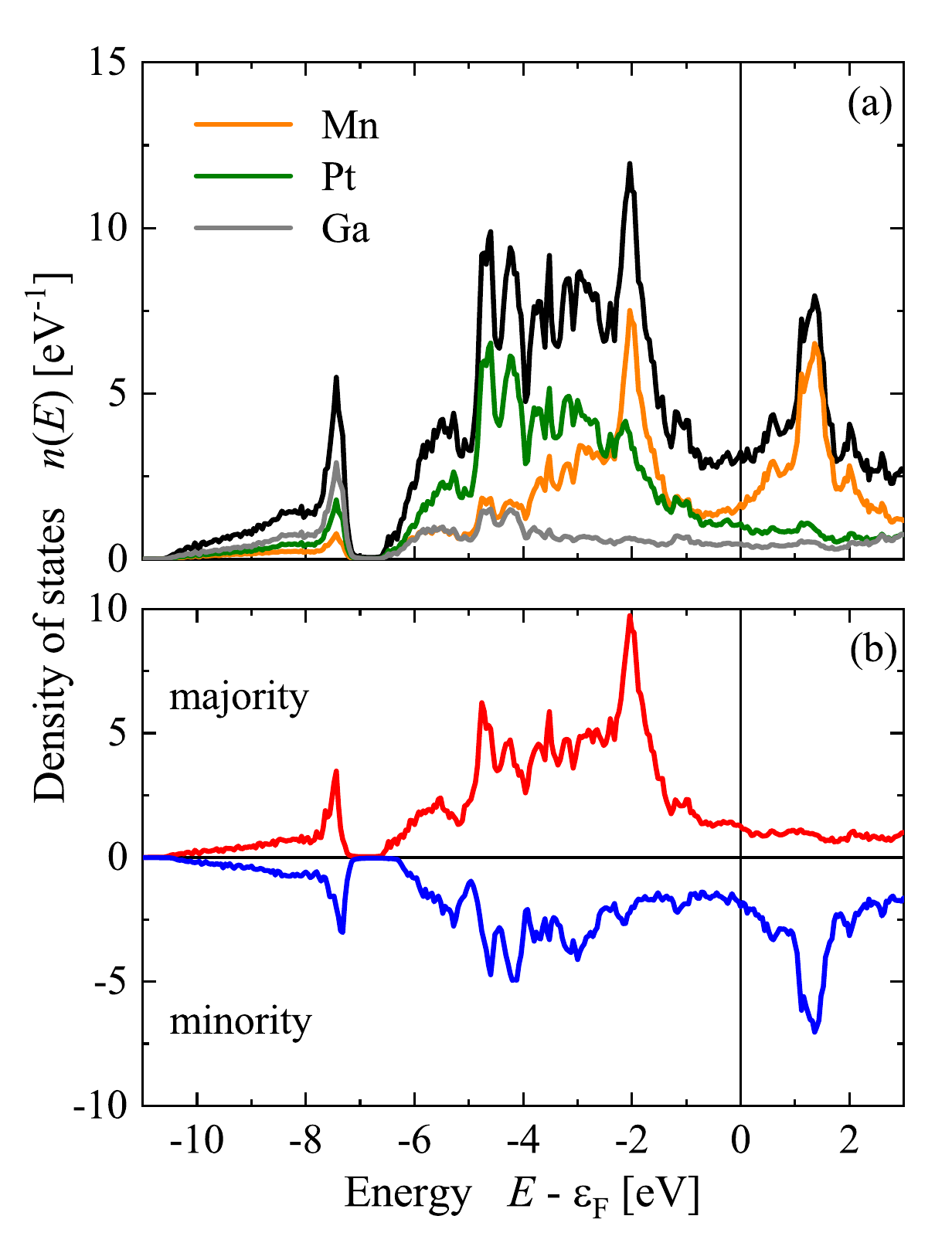}
   \caption{Electronic structure of MnPtGa in the canted magnetic state
            with $\theta=35^\circ$.\\
            a) shows the site resolved and b) the spin resolved density of
            states with respect to the primitive cell containing six atoms.}
   \label{fig:canted_dos}
\end{figure}

The electronic structure of MnPtGa in the canted magnetic state is illustrated
in Figure~\ref{fig:canted_dos} by means of the density of states. Displayed are
the atomic and the spin resolved contributions to the density of states
($n(E)$). The total spin magnetic moment amounts to $m_s=6.22\:\mu_B$, and the
total orbital moment is $m_l=0.08\:\mu_B$. Both are parallel to the $c$ axis.

It is found that the Ga atoms contribute mainly to the low lying states at
energies below -7~eV. Those states appear below a hybridization gap and are
mainly of $s$ character. It is also seen that the Pt $d$ states are almost
completely occupied and are located close to the bottom of the valence states
with high densities in the energy range from -2 to -5~eV. The comparison of the
site and spin resolved contributions reveals that the magnetic state is
dominated by the localized Mn $d$ states. These states exhibit an exchange
splitting of approximately 3.4~eV, when estimated from the associated maxima of
$n(E)$. The spin and orbital magnetic moments at the two Mn atoms are given by
the vectors
$\vec{m}^{\rm Mn}_s = ( \pm2.19,\:0.00,\:3.12)\:\mu_B$ and
$\vec{m}^{\rm Mn}_l = ( \pm0.014,\:0.00,\:0.02)\:\mu_B$, respectively.
The resulting overall, total (spin + orbital) magnetic moment at the Mn atoms is
$m^{\rm Mn}=3.84\:\mu_B$, which is only slightly higher than the values found
for the collinear ferromagnetic and antiferromagnetic states. The net
ferromagnetic moment, that is the magnetic moment along $c$, reported in
Reference~\cite{CBS20} ranges from 2.6 to 3.1~$\mu_B$ for temperatures
between 10 and 200~K, which is close to the calculated value for the canted
state.

\begin{figure}[htb]
   \centering
   \includegraphics[width=8cm]{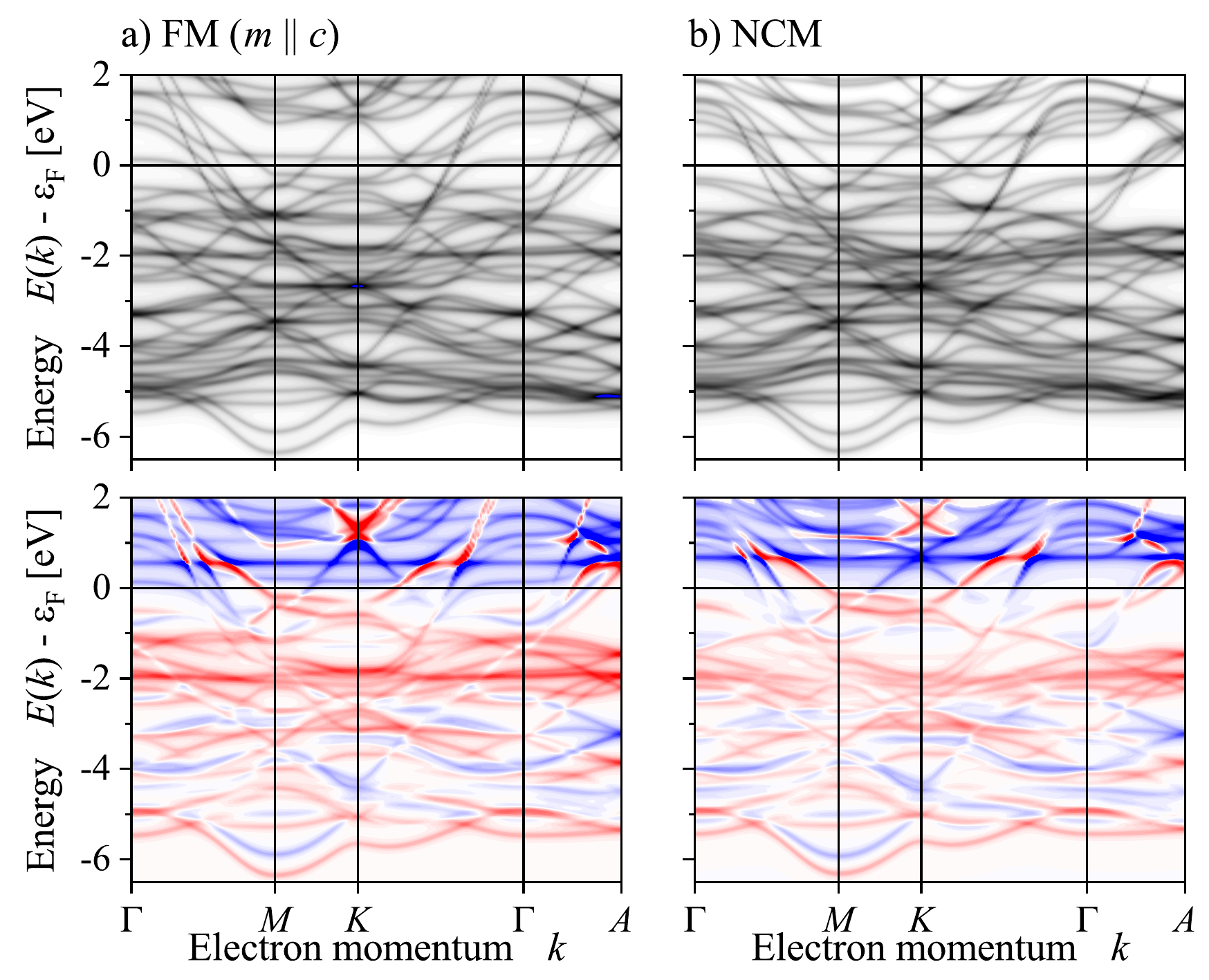}
   \caption{Comparison of the band structures of MnPtGa in the ferromagnetic
            (fm) state with $m$ along the $c$-axis (a) and non-collinear,
            canted magnetic state (ncm) with $\theta=35^\circ$ (b).\\
            Plotted are the total Bloch spectral function in gray and the spin
            resolved Bloch spectral functions with majority electron states in
            red and minority states in blue.}
   \label{fig:bsf}
\end{figure}

More details of the electronic structure of the canted state are shown in
Figure~\ref{fig:bsf} together with a comparison to the ferromagnetic state. The
band structures are very similar with some details shifted in energy or momentum
space. Differences in the obtained band crossings are due to the different
symmetries of the two cases that are the magnetic space group $P\:6_3/mm'c'$ in
the ferromagnetic and $P\:2_1/m.1$ in the canted magnetic state. Such
differences are easiest seen above the Fermi energy at the $K$ point of the band
structures. Various symmetry related differences are also observed for the band
positions and splitting at the $\Gamma$ point. However, the exchange splitting
-- which is seen at the bottom of the bands at the $M$ point -- remains nearly
unchanged (compare also Figure~\ref{fig:dlm-bsf}, where this splitting is absent
in the paramagnetic and antiferromagnetic states).

\section{Summary} 

From the calculation of collinear magnetic order it is found that the
equilibrium energy of the antiferromagnetic order is slightly lower compared to
ferromagnetic order. This gives a hint that the low temperature magnetic ground
state may be antiferromagnetic rather than ferromagnetic. The calculated Curie
temperature is  higher compared to the N{\`e}el temperature. Therefore, it can
be concluded that the high and low temperature magnetic order is different, and
a magnetic phase transition may appear.

Calculations for non-collinear magnetic states reveal spiral or canted spin
order. The spin spirals exhibit -- independent on the initial spin order -- the
lowest energies for $q$ along the [0001] direction. They give some hints on
frustrated magnetism as the energies do not change much over wide $q$ ranges.
The ferromagnetic cone spiral has a minimum for a cone angle close to
$\arctan(c/a)$. In both -- ferromagnetic and antiferromagnetic -- types of cone
spirals, the energy minimum appears at $q=0.5\pi/c$. However, various local
minima are observed for the ferromagnetic spirals, pointing at metastable
configurations. Therefore, a static canting of the magnetic moments is possible.
From the calculations for canted spin order, the lowest energy is found for a
canting angle of about $38^\circ$. Thus, the ferromagnetic (parallel aligned)
part of the magnetic moments is along the $c$ axis, whereas the
antiferromagnetic (antiparallel aligned) parts appear in successive Mn planes.

Overall it is found that the energies for the various states -- ferromagnetic,
antiferromagnetic, spiral, canted -- are all rather close together. This is in
well agreement with the different types of magnetic order reported in the
literature, in particular in Reference~\cite{CBS20}. Mostly, the energy
differences are in the order of not more than 30~meV, independent on the used
method. From this observation, it can be deduced that a particular magnetic state
may be forced when an external field is applied. This includes more complicated
non-collinear states as for example skyrmions or antiskyrmions. Considering the
effect of temperature, a frustrated state frozen in a spin glass might be changed
into a canted state already when a very small magnetic field is applied at
finite temperature.


\bigskip
\begin{acknowledgments}

This work is dedicated to Prof. emer. J{\"u}rgen K{\"u}bler who died in January
2025. The authors are grateful for his valuable suggestions and discussions.

\end{acknowledgments}

\section*{Data Availability Statement}

The data that support the findings of this study are available within the
article. Additional data are available from the corresponding author upon
reasonable request.

\bigskip
\appendix

\section{Magnetic anisotropy in hexagonal systems} 
\label{app:app1}

The general theory of the magneto-crystalline anisotropy is described in
textbooks~\cite{Get08,MJi12}. In most practical cases, only the uniaxial term of
the magnetic anisotropy is given for hexagonal or tetragonal systems. However,
this does not reflect the symmetry of the crystal structure. For hexagonal
crystals, the uniaxial anisotropy constant may be calculated from the energy
difference for magnetization along the $c$ axis and an in plane direction e.g.:
   $K_u = E^{10\overline{1}0}-E^{0001}$,
which is not unique because the in plane Cartesian orthogonal coordinates $x$
and $y$ are not equivalent in hexagonal systems. In the following terms up to
$6^{\rm th}$ order are investigated to account for the 6-fold rotational
symmetry of hexagonal systems. The extensions on higher order terms in the
angular dependence of the magnetic anisotropy made here are similar to those for
tetragonal systems reported in Reference~\cite{FHF21}.

Care must be taken about the directions has to be taken when calculating the
magneto-crystalline energy from the first principles energies using different
computer programs. For instance, the direction of magnetization (quantization
axis) may be given either in terms of lattice vectors or Cartesian vectors,
which differ in hexagonal systems. The three lattice vectors for the hexagonal
structures are defined by:
  $\vec{g_1} = (\sqrt{3},\:-1,\:0)a/2$,
  $\vec{g_2} = (0,\:1,\:0)a$, and
  $\vec{g_3} = (0,\:0,\:1)c$,
that is the vectors along $x$, $y$, and $z$ are given by the
combinations of lattice vectors $2\vec{g_1} + \vec{g_2}$, $\vec{g_2}$, and
$\vec{g_3}$, respectively. In other notation using $hkil$ as indices one has:

\begin{eqnarray}
 \vec{g}_{10\overline{1}1} & = & (\sqrt{3},\:-1,\:0)a/2 ,  \nonumber  \\
 \vec{g}_{01\overline{1}1} & = & (0,\:1,\:0)a,             \nonumber  \\
 \vec{g}_{0001}            & = & (0,\:0,\:1)c,
\end{eqnarray}

and the basis vectors in Cartesian co-ordinates are derived from:

\begin{eqnarray}
 \vec{x} & = & (2\vec{g}_{10\overline{1}1}+\vec{g}_{01\overline{1}1})/\sqrt{3} ,  \nonumber  \\
 \vec{y} & = & \vec{g}_{01\overline{1}1},             \nonumber  \\
 \vec{z} & = & \vec{g}_{0001},
\label{eq:xyz}
\end{eqnarray}

Using a series expansion up to the $6^{\rm th}$ order in $\sin(\theta)$, the
magneto-crystalline energy of a hexagonal system is expressed as:

\begin{equation}
  E^{\rm hex}_{\rm crys} = \sum_{\nu=0}^{3} K_\nu \sin^{2\nu}(\Theta) + K_4 \sin^6(\Theta)\cos(6\Phi)
\label{eq:Eaniso}
\end{equation}

In the following, the lower index "crys" will be omitted and the energies are
indexed only by directions or by "hex".  From the high symmetry directions
[$h,k,i,l$] one has for variation of $h$ or $k$ (where $i=-(h+k)$) at fixed
$l=1$:

\begin{eqnarray}
 E^{0001}            & = & K_0,                      \nonumber \\
 E^{10\overline{1}1} & = & \sum_{{\nu=0,3}} K_\nu \sin^{2\nu}(\Theta^{10\overline{1}1}) - K_4 \sin^{6}(\Theta^{10\overline{1}1}), \nonumber  \\
 E^{20\overline{2}1} & = & \sum_{{\nu=0,3}} K_\nu \sin^{2\nu}(\Theta^{20\overline{2}1}) - K_4 \sin^{6}(\Theta^{20\overline{2}1}), \nonumber  \\
 E^{02\overline{2}1} & = & \sum_{{\nu=0,3}} K_\nu \sin^{2\nu}(\Theta^{02\overline{2}1}) - K_4 \sin^{6}(\Theta^{02\overline{2}1}).
\label{eq:ehki1}
\end{eqnarray}

Instead of $E^{20\overline{2}1}$ and $E^{02\overline{2}1}$ one can use
$E^{01\overline{1}1}$ and $E^{21\overline{3}1}$, respectively.

\begin{eqnarray}
  E^{01\overline{1}1} & = & \sum_{{\nu=0,3}} K_\nu \sin^{2\nu}(\Theta^{01\overline{1}1}) - K_4 \sin^{6}(\Theta^{01\overline{1}1}), \nonumber  \\
 E^{21\overline{3}1} & = & \sum_{{\nu=0,3}} K_\nu \sin^{2\nu}(\Theta^{21\overline{3}1}) + K_4 \sin^{6}(\Theta^{21\overline{3}1}).
\label{eq:ehki1alt}
\end{eqnarray}

Using  $z=c/a$, the various angles are found from
$\Theta^{10\overline{1}1} = \Theta^{01\overline{1}1} = \arctan(1/z)$,
$\Theta^{20\overline{2}1} = \Theta^{02\overline{2}1} = \arctan(2/z)$, and
$\Theta^{21\overline{3}1} = \arctan(\sqrt{3}/(2z))$.
Note that the direction $[21\overline{3}1]$ is along the $x$ axis (see
Equation~(\ref{eq:xyz})), whereas $[01\overline{1}1]$ is along $y$.
Alternatively, $E^{10\overline{1}2}$ with
$\Theta^{10\overline{1}2} = \arctan(1/(2z))$ may be used (and similar by
increasing $l$ at fixed $h,k$).

For in-plane vectors with vanishing $z$ component one finds $\Theta=\pi/2$ such
that $\sin^n(\Theta)=1$ and thus for two perpendicular in-plane directions
because $\cos(n\pi)=1$ and $\cos((2n-1)\pi)=-1$ ($n\in\mathbb{N}_0$):

\begin{eqnarray}
 E^{10\overline{1}0} & = & \sum_{{i=0,3}} K_i - K_4,           \nonumber  \\
 E^{01\overline{1}0} & = & \sum_{{i=0,3}} K_i - K_4,           \nonumber  \\
 E^{11\overline{2}0} & = & \sum_{{i=0,3}} K_i - K_4,
\label{eq:ehki0}
\end{eqnarray}

keeping in mind that $\phi_{10\overline{1}0}=-\pi/6$ and
$\phi_{01\overline{1}0}=\pi/2$ or $\phi_{21\overline{3}0}=0$ and
$\phi_{12\overline{3}0}=\pi/6$. Further, it is found for example that $2
K_4=E^{12\overline{3}0}-E^{10\overline{1}0}$, or $E^{21\overline{3}0}-
E^{01\overline{1}0}$, etc..
Instead of $E^{11\overline{2}0}$ one can also use one of the following energy relations.

\begin{eqnarray}
 E^{1\overline{1}00} & = & \sum_{{i=0,3}} K_i + K_4,           \nonumber  \\
 E^{12\overline{3}0} & = & \sum_{{i=0,3}} K_i + K_4,           \nonumber  \\
 E^{21\overline{3}0} & = & \sum_{{i=0,3}} K_i + K_4
\label{eq:ehki0alt}
\end{eqnarray}

Finally, the anisotropy constants $K_i$ are found from the equations:

\begin{flalign}
 K_0 & = E^{0001}, &
\end{flalign}

\begin{flalign}
 K_1 = & -\frac{1}{12z^4}[(60 E^{0001} + 4 E^{10\overline{1}1} - 64 E^{10\overline{1}2}) z^6  \nonumber  \\
       & +(36 E^{0001} + 12 E^{10\overline{1}1} - 48 E^{10\overline{1}2}) z^4                 \nonumber  \\
       & +(12 E^{10\overline{1}1} - 12 E^{10\overline{1}2}) z^2                               \nonumber  \\
       & + 4 E^{10\overline{1}1} - 4 E^{10\overline{1}2}  - 3 E^{21\overline{3}0}]            &
\end{flalign}

\begin{flalign}
 K_2 = &  \frac{1}{12z^4}[(48 E^{0001} + 16 E^{10\overline{1}1} - 64 E^{10\overline{1}2}) z^8  \nonumber  \\
       & +(120 E^{0001} + 56 E^{10\overline{1}1} - 176 E^{10\overline{1}2}) z^6                \nonumber  \\
       & +(36 E^{0001} + 72 E^{10\overline{1}1} - 108 E^{10\overline{1}2}) z^4                 \nonumber  \\
       & +(40 E^{10\overline{1}1} - 25 E^{10\overline{1}2} - 15 E^{21\overline{3}0}) z^2       \nonumber  \\
       & + 8 E^{10\overline{1}1} - 2 E^{10\overline{1}2} - 6 E^{21\overline{3}0} ]             &
\end{flalign}

\begin{flalign}
 K_3 = & -\frac{1}{12z^4}[(48 E^{0001} + 16 E^{10\overline{1}1} - 64 E^{10\overline{1}2}) z^8     \nonumber  \\
       & +(60 E^{0001} + 52 E^{10\overline{1}1} - 112 E^{10\overline{1}2}) z^6                    \nonumber  \\
       & +(12 E^{0001} -  6 E^{10\overline{1}0} - 6 E^{21\overline{3}0} + 60 E^{10\overline{1}1} - 60 E^{10\overline{1}2}) z^4  \nonumber  \\
       & +(28 E^{10\overline{1}1} - 13 E^{10\overline{1}2} - 15 E^{21\overline{3}0}) z^2          \nonumber  \\
       & + 4 E^{10\overline{1}1} - E^{10\overline{1}2} - 3 E^{21\overline{3}0}]                   &
\end{flalign}

\begin{flalign}
 K_4 & = -\frac{1}{2} [E^{10\overline{1}0} - E^{21\overline{3}0}], &
\end{flalign}

where $z=c/a$. Further, equivalent energy terms from the above given equations
may be used alternatively (for examples of equivalent energies, see
Equations~(\ref{eq:ehki1alt}) and~(\ref{eq:ehki0alt})).

The magneto-crystalline anisotropy energy ($E_{a}$) is the difference between
the magneto-crystalline energy (here $E^{\rm hex})$ and the isotropic
contribution, that is, the spherical part $K_0$:

\begin{equation}
  E_{a} = E^{\rm hex} - K_0
\end{equation}

From this equation, the magneto-crystalline anisotropy energy may be positive or
negative, depending on the direction and values of the $K's$. This makes a
visualization by plotting the three dimensional distribution of
$E_a(\vec{r})=E_a(\theta,\phi)$ complicated. Therefore, the
alternative anisotropy energy $E_{a'}$ with respect to the lowest energy is
plotted most often, where:

\begin{equation}
  E_{a'} = E_{a} - \min(E_{a}) = E^{\rm hex} - \min(E^{\rm hex}),
\end{equation}

which is still positive although $E_a<0$.

\section{Magnetic anisotropy and applied field} 
\label{app:app2}

\begin{figure}[htb]
   \centering
   \includegraphics[width=8cm]{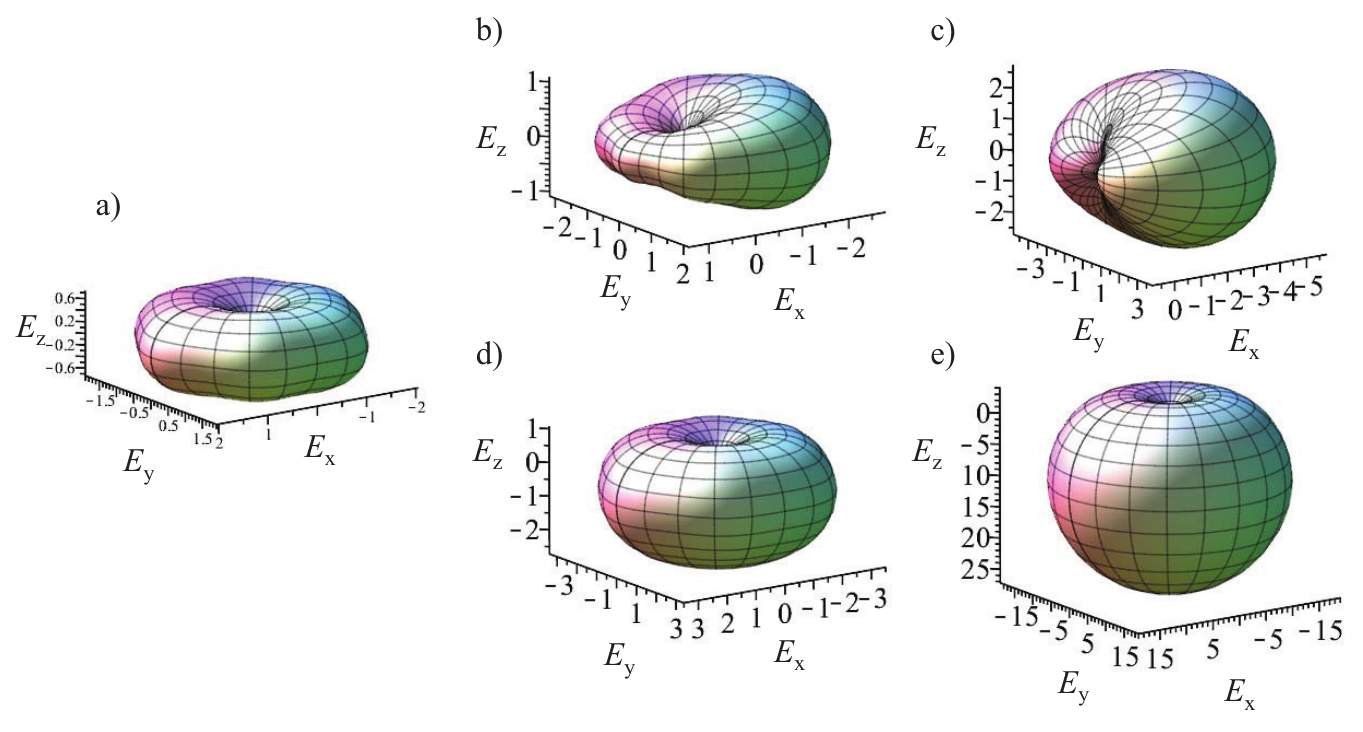}
   \caption{Influence of an applied field on the magnetic energy distribution
            in the ferromagnetic case.\\
            Shown are the distributions for a field along $x$ (b,c) or $z$ (d,e)
            for low (b,d) and high (c,e) fields compared to Zero (a) field.}
   \label{fig:field}
\end{figure}

The magnetic energy is altered if a magnetic field is applied. This has
consequences on the determination of the magnetic anisotropy. The magnetic
energy in an induction field $B_0=\mu_0H$ is~\cite{Get08,CGr09,MJi12}:

\begin{equation}
  E(\Theta,\Phi,\vec{H}) = E_{ani}(\Theta,\Phi)- M_sB_0 (\alpha_x\beta_x + \alpha_y\beta_y + \alpha_z\beta_z),
\end{equation}

where $\alpha_i=\alpha_i(\Theta,\Phi)$ and $\beta_i=\beta_i(\theta,\phi)$ are
the direction cosines of the magnetization and the applied field, respectively.
Here, $E_{ani}(\Theta,\Phi)$ corresponds to the 6$^{\rm th}$-order anisotropy
energy given by Equation~(\ref{eq:Eaniso}). The spontaneous magnetization
$M_s=M_s(T)$ depends on the temperature, as well as the anisotropy constants
$K_i$ do. The effect of the magnetic field on the magnetic energy distribution
is illustrated in Figure~\ref{fig:field} for the case of ferromagnetic order
(compare Figure~\ref{fig:aniso}~fm).

\section{Dipolar magnetic anisotropy} 
\label{app:dipani}

In non-cubic systems, the dipolar anisotropy does not vanish and also
contributes to the magnetocrystalline anisotropy (see:
References~\cite{AYo77,MCa79,FHF21}). The complete Ewald lattice sum to
find the dipolar energy reads:

\begin{equation}
\label{eq:dipaniso}
   E_{dip} = \frac{\mu_0}{8\pi}
             \sum_{i}^{N_{at}} \sum_{j}^{N_{at}} \sum_{n=h,k,l}^{N_{h,k,l}}
             \left[
             \frac{\vec{m}_i \cdot \vec{m}_j}{|R_{n}|^3}
                       - 3\frac{(\vec{R}_{n}\cdot \vec{m}_i)(\vec{R}_{n}\cdot \vec{m}_j)} {|R_{n}|^5}
             \right].
\end{equation}

where $N_{at}$ is the number of atoms in the cell and $\vec{R}_{n} =
\vec{r}_{ij} + \vec{R}_{hkl}$ and $\vec{r}_{ij}=\vec{r}_{i}-\vec{r}_{j}$ are the
vectors between all atoms used in the triple lattice summation. Further, on has
the important condition that $j \neq i$ for $\vec{R}_{hkl}=0$.
%
%
For hexagonal systems with lattice parameters $a$ and $c$ and the three lattice
vectors $g_i$ of Appendix~\ref{app:app1}, one has
$\vec{R}_{hkl} = h\vec{g}_1 + k\vec{g}_2 + l\vec{g}_3)$,
with $h,k,l \in \mathbb{Z}$.

The total number of lattice vectors is restricted by $\min(R_{\max},|R_{hkl}|)$
for a confining sphere with radius $R_{\max}$. For a given sphere, the upper
limits for the $h,k,l$ may be estimated by $N_{h,k}>R_{\max}/a$ and
$N_{l}>R_{\max}/c$. Alternatively, one may also restrict $R_{hkl}$ to stay
within a given cube. Wang and Holm~\cite{WHo01} estimated the cut-off error when
truncating the Ewald summation in Equation~(\ref{eq:dipaniso}).

In general, the individual magnetic moments $\vec{m}_{i,j}$ do not necessarily
have to be collinear. Their magnitude depends on the direction of magnetization
$\hat{n}=\vec{M}/|M|$ such that not only the dipolar energy
$E_{dip}=E_{dip}(\hat{n})$ but also the magnetic moments
$\vec{m}_{i,j}=\vec{m}_{i,j}(\hat{n})$ become directional dependent. To find the
different magnitudes of the magnetic moments, one has to perform calculations
that are fully relativistic or at least include the spin--orbit interaction as
perturbation.

Finally, the dipolar anisotropy is given by the difference between the energies
for two different magnetization directions:

\begin{equation}
     \Delta E_{\rm aniso}=E(\hat{n}_2)- E(\hat{n}_1).
\end{equation}

Two well-distinguished directions in hexagonal systems are $\vec{n}_1=[0001]$
and $\vec{n}_2=[01\overline{1}0]$, which are along the $c$ axis and in the basal
plane along $a$, respectively. Positive values indicate an easy dipolar
direction that is along the $[0001]$ axis. It has a second-order angular
dependence. Alternatively, a different in-plane direction, say
$\vec{n}=[11\overline{2}0]$, $[1\overline{1}00]$, or $[10\overline{1}0]$, may
also be used and the results for inequivalent directions should be compared.

In case that the magnetic moments are collinear and aligned along the
magnetization direction, Equation~(\ref{eq:dipaniso}) becomes:

\begin{equation}
\label{eq:dipanisodir}
   E_{dip}(\hat{n}) = \frac{\mu_0}{8\pi}
                      \sum_{i} \sum_{j} m_i m_j \sum_{n}
                      \left[
                      \frac{\hat{n} \cdot \hat{n}}{|R_{n}|^3}
                       - 3\frac{(\vec{R}_{n}\cdot \hat{n})^2} {|R_{n}|^5}
                      \right].
\end{equation}

The equation becomes further simplified because of $\hat{n} \cdot \hat{n}=1$.
The signs and values of the magnetic moments may still depend on the direction
of magnetization. The lattice vectors $\vec{R}_{n}(i,j,h,k,l)$ still depend on
the relative positions ($\vec{r}_{i}-\vec{r}_{j}$) of the atoms in the basic
cell. The last summation over $n$ depends only on the structure and is called
the magnetic dipolar Madelung sum $M_{ij}$:

\begin{equation}
\label{eq:Madelung}
   M_{ij}(\hat{n}) = \sum_{n}
                     \left[
                     \frac{1}{|R_{n}|^3}
                       - 3\frac{(\vec{R}_{n}\cdot \hat{n})^2} {|R_{n}|^5}
                     \right].
\end{equation}

The $j$ dependent part of Equation~(\ref{eq:dipanisodir}), is often introduced
as the magnetic dipolar field $\vec{B}$. The dipolar field $\vec{B}_i$ at each
atom of the cell is calculated again from a direct Ewald lattice sum:

\begin{eqnarray}
\label{eq:dipfield}
   \vec{B_{i}} (\hat{n}) & = & \frac{\mu_0}{4\pi} \sum_{j} \sum_{n}
                               \frac{m_j}{|R_{n}|^3}
                               \left[ 3 (\hat{n} \cdot \hat{r}_{n})\hat{r}_{n} - \hat{n} \right] \nonumber \\
                         & = & \frac{\mu_0}{4\pi} \sum_{j} m_j M_{ij}(\hat{n}),
\end{eqnarray}

where $\hat{r}_n= \vec{R}_n/|R_n|$ represents the direction of the distance
vectors. $\hat{r}_n$ depends on the position $r_i$ of each atom in the primitive
cell. Note, in some work the double sum $\sum_{i}\sum_{j \geq i}$ is used
instead of $1/2 \sum_{i}\sum_{j}$. The restriction on the sums are the same as
for Equation~(\ref{eq:dipaniso}).

Alternatively to (\ref{eq:dipaniso}), the dipolar energy may be calculated  from:

\begin{equation}
\label{eq:edipaniso}
   E_{dip} (\hat{n}) = - \frac{1}{2}\sum_{i} \vec{m}_i \cdot \vec{B}_i.
\end{equation}

\begin{figure}[htb]
   \centering
   \includegraphics[width=8cm]{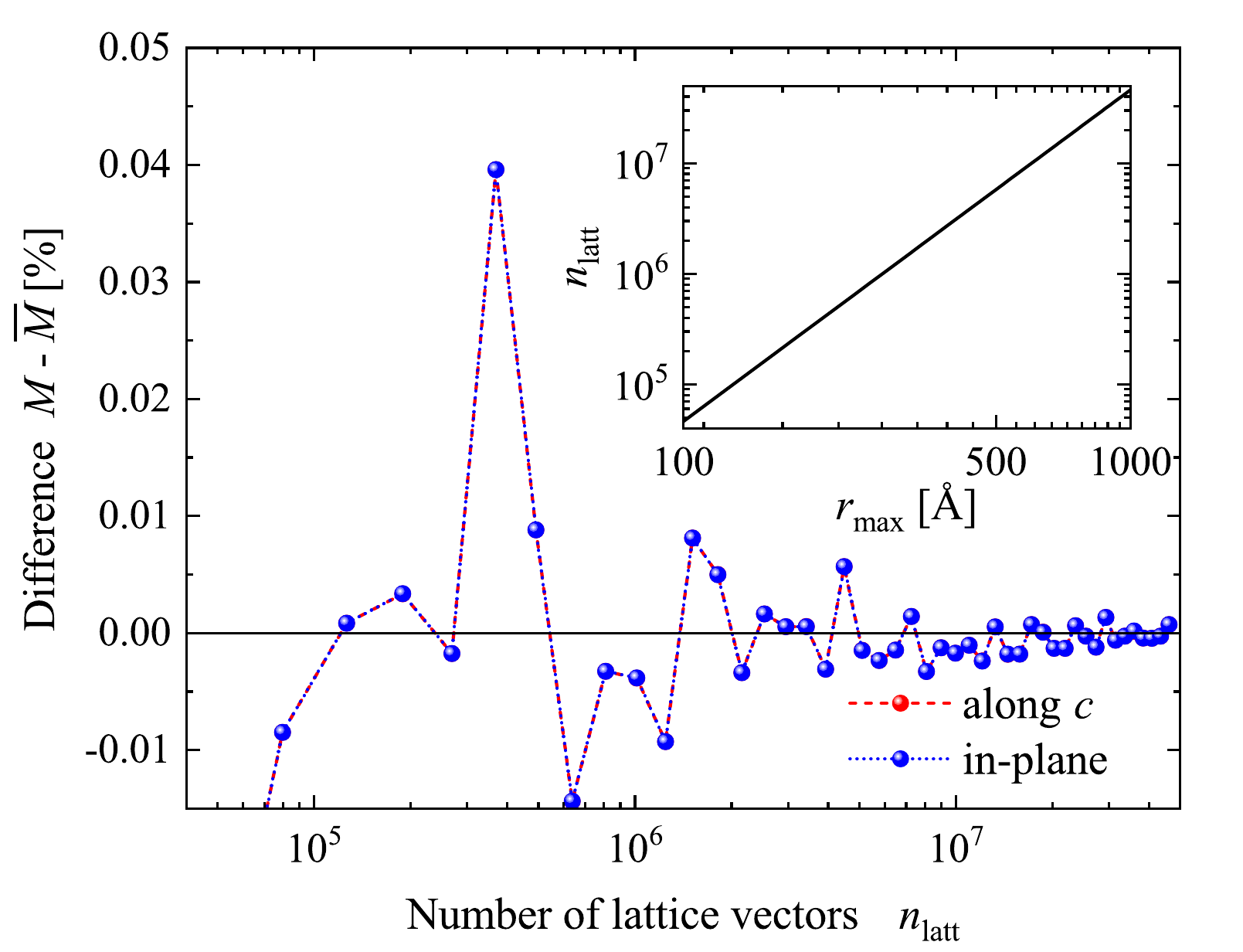}
   \caption{Relative deviation of the Madelung sum as function of the number
            of lattice vectors $n_{latt}$.\\
            Shown is the relative difference with increasing size of the
            confining sphere. The inset displays the dependence of $n_{latt}$ on
            the radius of the confining sphere. ($\overline{M}$ is the mean
            value for $n_{latt}>10^7$.) }
   \label{fig:madelung}
\end{figure}

Care must be taken that the Ewald lattice summations in the above equations
converge. Figure~\ref{fig:madelung} shows the behavior of the magnetic dipolar
Madelung sum $M_{ij}(\hat{n})$ of MnPtGa as a function of the radius $R_{\max}$ of
the confining sphere. More than 40~million lattice vectors corresponding to
$R_{\max}\geq 1000$~{\AA} are needed to reach convergence. In
the present work, $R_{\max}=1785$~{\AA} was used, corresponding to approximately
410 times the lattice parameter $a$. Note that the ferromagnetic dipolar
Madelung sum is negative for $\hat{n}$ along the $c$ axis.


\bigskip
\bibliography{MnPtGa}   

\end{document}